\documentclass{iopjournal}

\usepackage{amsmath}
\usepackage{amssymb}
\usepackage{dcolumn}
\usepackage{bm}
\usepackage{mathrsfs}
\usepackage[ruled,vlined,]{algorithm2e}

\begin{document}

\articletype{Paper} 

\title{Efficient Identification of Critical Regions via Flow Matching–Based Monte Carlo Initialization}

\author{Qian-Rui Lee$^{1,*}$\orcid{0009-0009-2250-2419}, and Daw-Wei Wang$^{1,2,3}$\orcid{0000-0001-8160-6570}}

\affil{$^1$Physics Department, National Tsing Hua University, Hsinchu 30013, Taiwan}

\affil{$^2$Center for Theory and Computation, National Tsing Hua University, Hsinchu 30013, Taiwan}

\affil{$^3$Center for Quantum Technology, National Tsing Hua University, Hsinchu 30013, Taiwan}

\affil{$^*$Author to whom any correspondence should be addressed.}

\email{liqianrui369@gmail.com}


\begin{abstract}

Markov chain Monte Carlo (MCMC) is a standard tool for studying many-body systems, but its practical cost can become substantial, especially when simulations must be repeated across temperatures and lattice sizes or near transition regions where equilibration becomes increasingly difficult. In this work, we introduce a Flow Matching (FM) framework. It is not a standalone replacement for equilibrium Monte Carlo. Instead, we present it as a scalable, physically informed initializer for downstream MCMC simulations. We use a U-Net architecture. The FM model is trained on small-system configurations of the 2D XY model and then deployed across unseen temperatures and larger lattice sizes.
FM-generated configurations preserve the correct qualitative physical trends across temperature and system size. This makes them suitable warm-start states for subsequent Monte Carlo refinement. Observables computed from FM-generated samples primarily serve as diagnostics of initializer quality, not as precision equilibrium estimates. The regression-based $L_2$ objective suppresses variance and limits the accuracy of fluctuation-sensitive observables. Examples include susceptibility and spin stiffness. Still, the model captures sufficient local statistical structure to yield physically aligned initial states across a broad range of conditions.
These results support a reusable hybrid FM--MCMC workflow. The one-time FM training cost can be amortized across temperatures and lattice sizes. The generated warm-start configurations then reduce the burden of initializing large-scale Monte Carlo simulations. Our findings show that Flow Matching can support efficient exploration of transition regions in many-body systems by providing reusable warm-start configurations for downstream Monte Carlo simulations.

\end{abstract}

\section{Introduction}
Monte Carlo—especially Markov chain Monte Carlo (MCMC)—is one of the most powerful computational tools across three fronts: in condensed matter, it is the workhorse for probing phase transitions and complex systems~\cite{Landau_Binder_2005}; in quantum many-body physics, quantum Monte Carlo is among the few reliable non-perturbative methods~\cite{Sandvik_2010, RevModPhys.73.33}; and in lattice QCD, Euclidean-lattice Monte Carlo provides the only systematically improvable first-principles non-perturbative approach at low energies~\cite{RevModPhys.82.1349, PhysRevLett.49.613}. Algorithmically, MCMC targets the Boltzmann measure via detailed balance and ergodicity (e.g. Metropolis, HMC~\cite{HMC}, cluster~\cite{PhysRevLett_Wolff}), with uncertainties governed by the integrated autocorrelation time~\cite{Landau_Binder_2005}.

However, MCMC encounters critical slowing down as correlation lengths grow near continuous phase transitions and toward the continuum limit; in lattice QCD, topological modes can freeze, inflating autocorrelations and complicating error estimates~\cite{Schaefer_2011}. Fermionic models further face the sign problem, whose generic resolution is NP-hard~\cite{PhysRevLett.94.170201}. Progress relies on advanced updates~\cite{PhysRevLett_Wolff}, tempering/multilevel schemes~\cite{Hukushima_1996}, and ML-assisted samplers; notably, flow-based generative sampling methods have shown efficiency gains and promising proposal mechanisms in lattice field theories~\cite{PhysRevD_Flow_based_MCMC, kanwar2024flow_based_sampling_lattice_field, abbott2024practical_app_flows, PhysRevD_NF_MC_sampling, PhysRevD.106.014514}. 

\begin{figure}[!ht]
    \centering
    \includegraphics[width=0.5\textwidth]{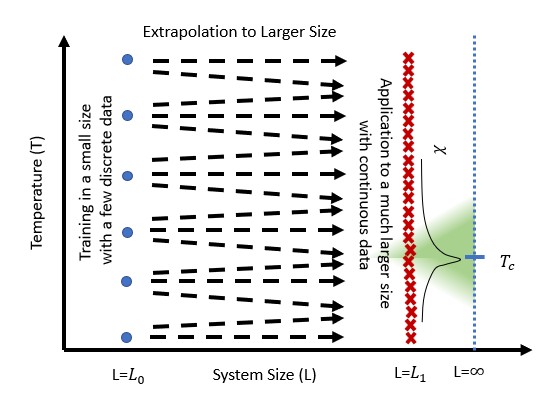}
    \caption{Schematic diagram of our work. Using Flow Matching with U-Net~\cite{ronneberger2015unet} inductive biases, we learn a deterministic probability-flow sampler that—trained only on $L_0\times L_0$ lattices—simultaneously interpolates across temperatures and extrapolates (dashed lines) to physically consistent configurations of $L_1\times L_1$ lattices with $L_1\gg L_0$, capturing reusable local statistical regularities and enabling scalable generation of physically informed warm-start configurations across temperatures and lattice sizes, which can be used to initialize downstream Monte Carlo refinement near the transition region. Here $\chi$ denotes the susceptibility, and the green shadow represents the critical regime.}
    \label{fig:schematic}
\end{figure}

Early successes with models like Normalizing Flows (NFs) demonstrated the potential to eliminate critical slowing down for a single set of system parameters~\cite{PhysRevD_Flow_based_MCMC, cheng2025lecture_notes_normalizing_flows, hackett2025_flow_based_sampling_multimodal_extendedmode}. However, the inefficiency of these initial approaches in exploring parameter space is a significant limitation. Studying critical phenomena requires simulations across various parameters (e.g., temperatures), but training a separate, computationally expensive model for each point is impractical~\cite{PhysRevD_NF_MC_sampling}. A key advancement to address this has been the development of conditional models, which learn the system's behavior across a continuous parameter range from sparse training data, enabling efficient interpolation and extrapolation~\cite{PhysRevD_NF_MC_sampling, Gerdes_2023LQF_equi_flow}. These models are often implemented with either architecturally rigid NFs, which can suffer from training instabilities like mode collapse~\cite{hackett2025_flow_based_sampling_multimodal_extendedmode, PhysRevDmode_collapse_for_flow_based}, or with Diffusion Models (DM)~\cite{ddpm, zhang2025efficient_unbiased_sampling_boltzmann, Wang2024_Diffusion_models_as_stochastic_quantization, Fukushima_2025_SQ_DM, zhu2024diffusionmodelslatticegauge}, which, while robust, typically rely on a computationally intensive iterative sampling process. 

Even as conditional models begin to address the challenge of exploring parameter space, a more fundamental and computationally demanding limitation persists in both traditional MCMC and these new generative approaches: the lack of generalizability across system sizes. Systematic studies of critical phenomena and the thermodynamic limit hinge on a sequence of simulations on progressively larger lattices. Under the current paradigm, each lattice size requires an entirely new, independent simulation or model training, a process whose computational cost scales prohibitively. Therefore, a truly efficient framework would not only interpolate across parameters like temperature but also capture the local, scale-independent correlation patterns, enabling extrapolation from a single, small-system training process to much larger lattices. The present work is designed to address this gap from the perspective of reusable initialization across temperatures and system sizes.

In this work we introduce a scalable generative framework in which Flow Matching (FM)~\cite{lipman2023flow} serves as a \emph{physically informed initializer} for Markov chain Monte Carlo (MCMC) simulations. Using a U-Net-based~\cite{ronneberger2015unet} FM model trained only on $32\times32$ configurations, we investigate whether the learned generator can be reused across temperatures and deployed on larger lattices to provide warm-start configurations for downstream Monte Carlo refinement.
Our emphasis, therefore, is not on treating FM-generated samples as precision equilibrium ensembles in their own right. Rather, the central question is whether they preserve sufficient of the correct local statistical structure and qualitative thermodynamic trends to efficiently initialize subsequent MCMC. We find that, although the regression-based $L_2$ objective suppresses ensemble variance and limits the accuracy of fluctuation-sensitive observables such as susceptibility and spin stiffness, the generated configurations remain physically well aligned at the qualitative level. This workflow is especially useful near transition regions, where physically informed initialization can reduce the cost of repeated Monte Carlo exploration across temperatures and lattice sizes.

The paper is organized as follows. We begin by introducing the theoretical framework of Flow Matching in Sec.~\ref{sec:theory_of_FM} and our training setup in Sec.~\ref{sec:Simulation_Framework}. 
Our main numerical results and analyses, covering qualitative diagnostics of FM-generated warm starts, cross-size transfer, error profiles of FM-only observables, and the amortized computational cost of reusable initialization, are presented in Secs.~\ref{sec:results_Increase_Sampling_Density} to \ref{sec:results_Time_Cost_Comparison}. 
We then provide a physical interpretation for these capabilities in Sec.~\ref{sec:disscuss_System_Size_Extrapolation}. 
The paper concludes with a discussion of our findings in Sec.~\ref{sec:discussion} and an outlook for future work in Sec.~\ref{sec:conclusion_and_outlook}. Further implementation details, more comprehensive results, and the source code for this study are included in the Appendices.

\section{Theoretical Structure of Flow Matching}
\label{sec:theory_of_FM}
Flow Matching bridges continuous normalizing flows and diffusion models by directly learning vector fields (flows) along predefined conditional probability paths between a simple base distribution (e.g., Gaussian noise) and the target data distribution. It trains models by regressing on these conditional velocities rather than relying on stochastic simulations, resulting in stable, scalable training and efficient sample generation via off-the-shelf ODE solvers.

\begin{figure*}[!t]
    \centering
    \includegraphics[width=\textwidth]{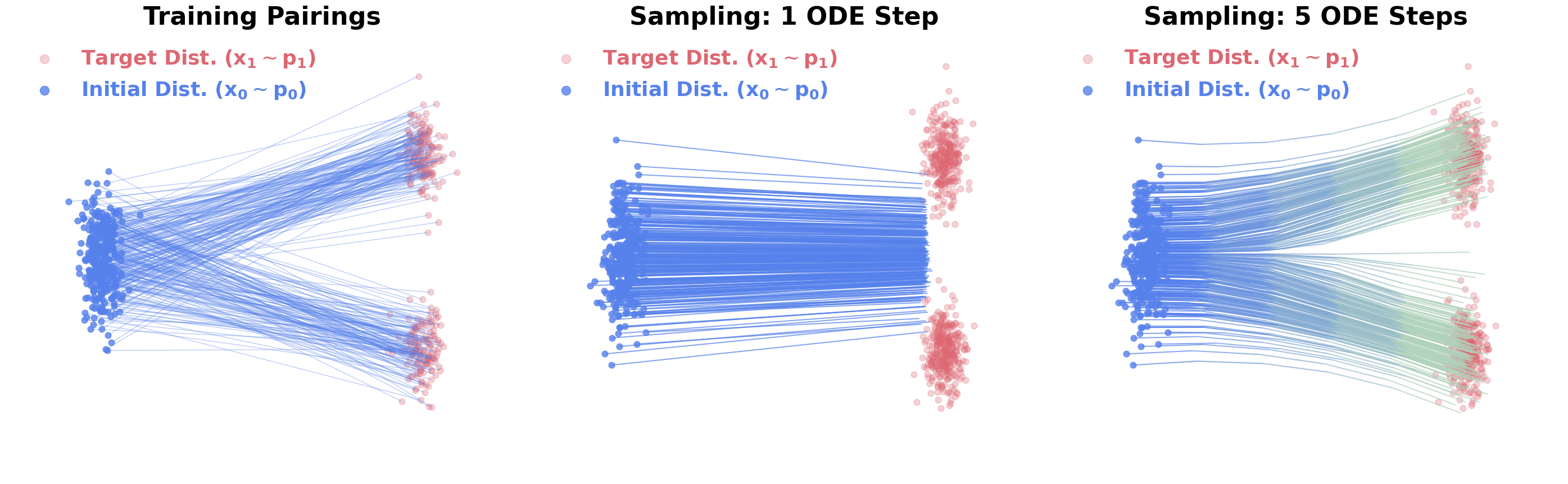}
    \caption{
    A naive Flow Matching example. When Flow Matching is trained as a deterministic ODE, trajectories from $\mathbf{v_{\theta}}$ deviate from the straight interpolation in Eq.~\eqref{eq_interpolation_path}. This is because $\mathbf{ v_\theta}$ learns the expectation of the conditional velocity, $\mathbf{v_t}(\mathbf{x}_t|\phi)$, induced by the random pairings of $(\mathbf{x}_0,\mathbf{x}_1)$ during training (left panel). Consequently, the first ODE iteration step (middle panel) points toward the target mean; with a finite number of iterations, inference preferentially produces samples near this mean, thereby underrepresenting the target variance and fluctuations.} 
    \label{fig:naive_flow_matching}
\end{figure*}

\subsection{Probability Flow ODE}
\label{sec:PFODE}
 Flow Matching admits arbitrary paths, granting flexibility to optimize sampling speed or modeling capacity~\cite{lipman2023flow}. In this work, we adopt a simple linear interpolation between base and data for simplicity. For a target data sample $\mathbf{x}_1 \sim \mathcal{P}_{data}$ and a standard Gaussian noise sample $\mathbf{x}_0 \sim \mathcal{N}(0, \mathbb{I})$ of the same dimensionality, we define the path $\{\mathbf{x}_t\}_{t=0}^{1}$ , where $t$ is a virtual time parameter indexing the continuous evolution of the distribution. The path at each subscripted time $t$ is given by

\begin{align}
    \mathbf{x}_t = t\; \mathbf{x}_1 + (1-t)\;\mathbf{x}_0
    \label{eq_interpolation_path}
\end{align}

\noindent For a fixed coupling $(\mathbf{x}_0, \mathbf{x}_1)$, the conditional velocity field is

\begin{align}
    \mathbf{v_t}(\mathbf{x}_t|\mathbf{x}_1) = \frac{d\mathbf{x}_t}{dt} = \mathbf{x}_1 - \mathbf{x}_0
    \label{eq_conditional_v}
\end{align}

\noindent But since $(\mathbf{x}_0, \mathbf{x}_1)$ are random, $\mathbf{v_t}(\mathbf{x}_t|\mathbf{x}_1)$ is also random. The true velocity field $\mathbf{u_t}(x)$ is an average of multiple conditional velocity fields. To know this average, consider a distribution $p_t(x)$ which satisfies the following boundary conditions.

\begin{align}
    p_{t=1}(\mathbf{x}_{t=1})=\mathcal{P}_{data}\quad and\quad p_{t=0}(\mathbf{x}_{t=0})=\mathcal{N}(0, \mathbb{I}).
    \label{eq_boundary_cond}
\end{align}

\noindent For simplicity, we can denote the configuration of all conditions as the variable $\phi$, which satisfies the distribution $q(\phi)$. The distribution of the probability path $p_t(\mathbf{x}_t)$ can be written as 

\begin{align}
    p_t(\mathbf{x}_t) = \int \left[\mathcal{D}\phi\right]\; p_t(\mathbf{x}_t|\phi)\; q(\phi)
    \label{eq_p_t}
\end{align}

\noindent Here, $p_t(\mathbf{x}_t|\phi)$ is the conditional probability path. For example, suppose that $\mathcal{P}_{data}$ is the Boltzmann distribution $exp(-\beta H)/Z$. The configuration $\phi$ of conditions is only the spin configurations $\mathbf{x}_1$ (generally, the condition can include temperatures, coupling constants, and other information that may affect the target distribution). Consider the path $\mathbf{x}_t$ in Eq.~\eqref{eq_interpolation_path},  because we choose $\mathbf{x}_0$ that satisfies the standard Gaussian distribution, therefore,  conditional probability $p_t(\mathbf{x}_t|\mathbf{x}_1)$  is the Gaussian distribution with mean $t\mathbf{x}_1$ and variance $(1-t)^2$~\cite{lipman2024flow_matching_guide_code}. The $p_t(\mathbf{x}_t)$ hence reads as

\begin{align}
    p_t(\mathbf{x}_t) = \int \left[\mathcal{D}\mathbf{x}_1\right]\;
    \mathcal{N}\left(\mathbf{x}_t;t \mathbf{x}_1, (1-t)^2\mathbb{I}\right)\; 
    \frac{e^{-\beta H(\mathbf{x}_1)}}{Z}
    \label{eq_p_t_ex}
\end{align}

\noindent We can easily check that Eq.~\eqref{eq_p_t_ex} satisfies the requirements in Eq.~\eqref{eq_boundary_cond}.


With conditional velocity field $\mathbf{v_t}(\mathbf{x}_t|\phi)$ and probability density $p_t(\mathbf{x}_t)$ in Eq.~\eqref{eq_p_t}, the continuity equation gives 

\begin{align}
    \frac{\partial\;p_t(\mathbf{x}_t)}{\partial t} 
    = - \nabla\cdot\int \left[\mathcal{D}\phi\right]\; \mathbf{v_t}(\mathbf{x}_t|\phi)p_t(\mathbf{x}_t|\phi)\; q(\phi)
    \label{eq_continuity}
\end{align}

\noindent Thus, with Bayes' theorem, the velocity field $\mathbf{u_t}(\mathbf{x}_t)$ is 

\begin{align}
    \mathbf{u_t}(\mathbf{x}_t) 
    = \int \left[\mathcal{D}\phi\right]\;
    \mathbf{v_t}(\mathbf{x}_t|\phi)\frac{p_t(\mathbf{x}_t|\phi)\; q(\phi)}{p_t(\mathbf{x}_t)} 
    = \int \left[\mathcal{D}\phi\right]\; \mathbf{v_t}(\mathbf{x}_t|\phi)\;
    p_t(\phi|\mathbf{x}_t)
\end{align}

\noindent The corresponding ODE to $\mathbf{u_t}(\mathbf{x}_t)$ is also called the probability flow ODE~\cite{song2021score_matching}:

\begin{align}
    \frac{d \mathbf{x}_t}{dt} = \mathbf{u_t}(\mathbf{x}_t) 
    \label{eq_pfode}
\end{align}

\noindent And we want the neural network $\mathbf{v_{\theta}}$ with learnable parameters $\theta$ to learn this velocity field $\mathbf{u_t}(\mathbf{x}_t)$. There are several methods for using neural networks to predict velocity fields. One reliable method is to let the neural network $\mathbf{net_{\theta}}$ predict the target sample and calculate the corresponding velocity field at different time steps~\cite{JIT}:

\begin{align}
    \mathbf{v_{\theta}}(\mathbf{x}_t, t) \equiv \frac{\mathbf{net_{\theta}}(\mathbf{x}_t, t)-\mathbf{x}_t}{1-t}
\end{align}

\noindent This expression follows directly from Eq.~\eqref{eq_interpolation_path} and Eq.~\eqref{eq_conditional_v}.

\subsection{Loss Function}
\label{sec:loss_fn}
With the probability flow ODE in Eq.~\eqref{eq_pfode}, the Flow Matching loss $\mathcal{L}_{FM}(\theta)$ can be defined as

\begin{align}
    \mathcal{L}_{FM}(\theta) \equiv 
    \int_{0}^{1}dt \int \left[\mathcal{D}\mathbf{x}_t\right] p_t(\mathbf{x}_t)
    \left[\mathbf{v_{\theta}}(\mathbf{x}_t,t) - \mathbf{u_t}(\mathbf{x}_t)\right]^2
    \label{eq_loss_fn_0}
\end{align}

\noindent However, we do not know the exact velocity field $\mathbf{u_t}(\mathbf{x}_t)$ from the training data. Thus, use the trick that the gradient of the loss function $\mathcal{L}(\theta)$ to update the neural network $\mathbf{v_{\theta}}$ is only depend on learnable parameters $\theta$, it is equivalent~\cite{lipman2023flow} to define conditional Flow Matching loss $\mathcal{L}_{CFM}(\theta)$ that we want to minimize as

\begin{align}
    \mathcal{L}_{CFM}(\theta) \equiv 
    \int_{0}^{1}dt \int \left[\mathcal{D}\mathbf{x}_0\mathcal{D}\phi\right]
    p_t(\mathbf{x}_t|\phi)q(\phi)
    \left[\mathbf{v_{\theta}}(\mathbf{x}_t,t|\phi) - \mathbf{v_t}(\mathbf{x}_t|\phi)\right]^2 
    \label{eq_loss_fn_1}
\end{align}

\noindent The loss function is similar to the action of the noninteracting fluid system with integration in the entire condition configuration $\phi$ and the initial Gaussian noise. For the path in Eq.~\eqref{eq_interpolation_path}, conditional velocity field $\mathbf{v_t}(\mathbf{x}_t|\phi)$ here is just $(\mathbf{x}_1 - \mathbf{x}_0)$ and see the Appendix~\ref{app:details_of_loss_fn} for more details.

\subsection{Flow of Random Sampling}
\label{sec:Flow_of_Random_Sampling}

Although the path in Eq.~\eqref{eq_interpolation_path} is straight, the training trajectories are created by randomly pairing data $\mathbf{x}_1$ and noise $\mathbf{x}_0$ under different conditions. Therefore, different conditional trajectories passing through the same $x_t$ at time $t$ generally have different directions, and the velocity field $\mathbf{u_t}(\mathbf{x}_t)$ is not identical to the constant vector $(\mathbf{x}_1 - \mathbf{x}_0)$ (see Fig.~\ref{fig:naive_flow_matching}). What the neural network $\mathbf{v_{\theta}}(\mathbf{x}_t,t|\phi)$ learned is an expectation over conditional velocities, and there is inherent uncertainty about the direction. One way to understand this is to view the training path as a quantum process, where a sample from the initial distribution $\mathbf{x}_0$ has the opportunity to reach any target distribution $\mathbf{x}_1$. Flow matching allows the neural network to learn the generators corresponding to the path that contributes most, the classical path. Due to this reason, instead of collapsing sampling to a single jump from noise to data, we must integrate the learned velocity field over a few time steps (however, research on the one-step sampling is rapidly expanding~\cite{geng2025meanflowsonestepgenerative, frans2025stepdiffusionshortcutmodels, deng2026generativemodelingdrifting}). Concretely, we divide the interval $t \in [0,1]$ into $N$ steps $\{t_k\}^N_{k=0}$ and numerically solve the problem (see Fig.~\ref{fig:naive_flow_matching}).

\begin{align}
    \frac{d\mathbf{x}_t}{dt} = \mathbf{v_{\theta}}(\mathbf{x}_t,t|\phi)
\end{align}

\noindent using a method such as explicit Euler or a higher‐order Runge–Kutta. Then, we can transport the samples from the distribution $p_0$ to $p_1$. For more training and sampling details, see the Appendix~\ref{app:train_and_sampling_strategy}.

\section{Generative Framework and Observable Estimation}
\label{sec:Simulation_Framework}

\subsection{Parameter-Wide Sampling via a Single Network}
\label{sec:handle_all_parameters}
One of the main advantages of flow-based sampling is that it can generate target samples in parallel and independently, thereby avoiding the Markov-chain autocorrelations that arise in traditional MCMC sampling. This advantage can be easily maximized in the Flow Matching training framework. To construct the conditional velocity field in Eq.~\eqref{eq_conditional_v}, only two types of information are required as conditions ($\phi$): one is the samples ($\mathbf{x}_1$) that obey the target distribution, and the other are the parameters (e.g., temperature in this work) corresponding to the samples. The Flow Matching framework can construct probability flows in completely parallel across the above two data dimensions. For example, the target samples $\mathbf{x}_1$ in this work have two indices: temperature and ensemble. Assuming that we have data corresponding to $N_T$ temperatures within a range, and each temperature has $N_{cf}$ spin configurations, the target sample $\mathbf{x}_1$, as a data array, has the shape ($N_T$, $N_{cf}$, $L$, $L$), where $L$ is the length and width of the lattice. Accordingly, the trained network $\mathbf{v_\theta}(\mathbf{x}_t,t,T)$ can generate, in parallel, arbitrarily large ensembles of independent $L\times L$ samples across the training temperature range.

\subsection{2D XY Model and Physical Observables}
\label{sec:Physical_Model_and_Observables}
To benchmark our framework, we use the standard two-dimensional (2D) XY model, defined on a square lattice of size $N = L \times L$. The system is described by the Hamiltonian:
\begin{align}
    H = -\sum_{\langle i,j \rangle} \cos(\Theta_i - \Theta_j)
\end{align}
where $\Theta_i \in (-\pi, \pi]$ is the angle of the spin at site $i$, and the sum runs over all nearest-neighbor pairs.

The canonical average of any thermodynamic observable $O$ is calculated over an ensemble of configurations:
\begin{align}
    \langle O \rangle = \frac{1}{N_{\text{cf}}} \sum_{k=1}^{N_{\text{cf}}} O_k
    \label{eq:ensemble_average}
\end{align}
where $O_k$ is the value of the observable for the $k$-th configuration and $N_{\text{cf}}$ is the total number of configurations in the ensemble. All intensive quantities are averaged over the $N$ lattice sites and the ensemble, as defined in Eq.~\eqref{eq:ensemble_average}. Throughout this work, the temperature $T$ is expressed in units where the Boltzmann constant $k_{\mathrm{B}}=1$.

The energy per site is derived directly from the Hamiltonian:
\begin{align}
    E \equiv \frac{1}{N} \langle H \rangle
\end{align}
The Cartesian components of the total magnetization are:
\begin{align}
    M_x \equiv \sum_{i=1}^{N} \cos(\Theta_i), \quad M_y \equiv \sum_{i=1}^{N} \sin(\Theta_i)
\end{align}
and the root-mean-square (RMS) magnetization per site is:
\begin{align}
    m \equiv \frac{1}{N} \left\langle \sqrt{M_x^2 + M_y^2} \right\rangle
\end{align}
The zero-field susceptibility is given by the fluctuations of the magnetization:
\begin{align}
    \chi \equiv \frac{1}{N T}
    \left(
    \left\langle M_x^2 + M_y^2 \right\rangle
    -
    \left\langle \sqrt{M_x^2 + M_y^2} \right\rangle^2
    \right)
\end{align}
To define the spin stiffness, we consider the negative Hamiltonian and the spin current restricted to bonds crossing the lattice in the $y$-direction:
\begin{align}
    h_y \equiv \sum_{\langle i,j \rangle_y} \cos(\Theta_i - \Theta_j), \quad
    I_y \equiv \sum_{\langle i,j \rangle_y} \sin(\Theta_i - \Theta_j)
\end{align}
The spin stiffness is then defined as~\cite{Sandvik_2010}:
\begin{align}
    \rho_s \equiv \frac{1}{N}
    \left(
    \langle h_y \rangle - \frac{1}{T} \langle I_y^2 \rangle
    \right)
\end{align}
Finally, the vortex density is obtained by summing the absolute winding number over each elementary $1\times 1$ plaquette ($\square$):
\begin{align}
    \rho_v \equiv \frac{1}{N}
    \left\langle
    \sum_{\square}
    \left|
    \frac{1}{2\pi} \sum_{\langle i,j \rangle \in \partial\square} [\Theta_j - \Theta_i]_{2\pi}
    \right|
    \right\rangle
\end{align}
where the inner sum is over the four bonds forming the plaquette's boundary $\partial\square$, and $[\cdots]_{2\pi}$ wraps the angle difference into the interval $(-\pi, \pi]$. 

The training data, consisting of spin configurations $\mathbf{x}_1$, uses angles mapped to the range $[0, 2\pi)$. The neural network is trained to directly output continuous angular values without an explicit modular arithmetic operation (i.e., no wrapping into $(-\pi, \pi]$ or $[0, 2\pi)$). This is a valid approach because all relevant physical observables, such as the energy and magnetization components, are calculated using trigonometric functions (e.g., $\cos(\Theta_i - \Theta_j)$ and $\sin(\Theta_i)$), which are inherently periodic with a period of $2\pi$. The full implementation of our model and the source code for all simulations are made publicly available to ensure reproducibility (see Sec. Data availability).

In the present work, these observables play two different roles. Observables computed from equilibrated MCMC ensembles serve as the physical reference, whereas the same observables computed from FM-generated samples are used primarily as diagnostics of whether the generated configurations are suitable warm starts, rather than as precision equilibrium estimates.

\section{Result I: Temperature-Interpolation Diagnostics for FM-Generated Warm Starts}
\label{sec:results_Increase_Sampling_Density}

\begin{figure}[!ht]
    \centering
        \includegraphics[width=\textwidth]{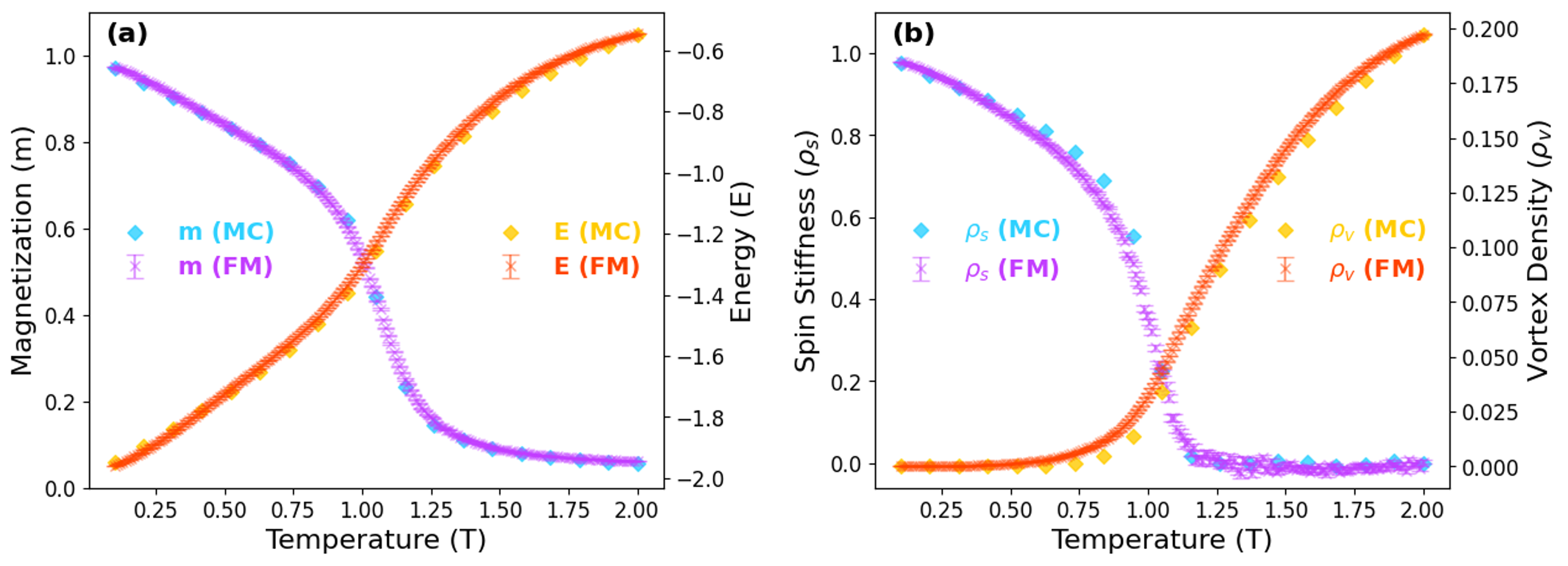}
        \caption{Thermodynamic observables for the $32\times32$ XY model, demonstrating the Flow Matching (FM) model's ability to increase sampling density via interpolation. The panels show (a) energy per site $E(T)$ and magnetization $m(T)$, and (b) spin stiffness $\rho_s(T)$ and vortex density $\rho_v(T)$. The sparse points represent MCMC (MC) data, which serves as the training set, sampled at a coarse temperature interval of $\Delta T = 0.10$. Each of these points was generated from an ensemble of $5000$ configurations. The denser set of points shows the results from FM, evaluated on a tenfold denser grid ($\Delta T = 0.01$), confirming agreement with and smooth interpolation between the training points.
        }
        \label{fig:eng_mag_stiffness_vortex_increase}
\end{figure}

\begin{figure*}[!t]
    \centering
    \begin{minipage}[b]{\textwidth}
        \includegraphics[width=\textwidth]{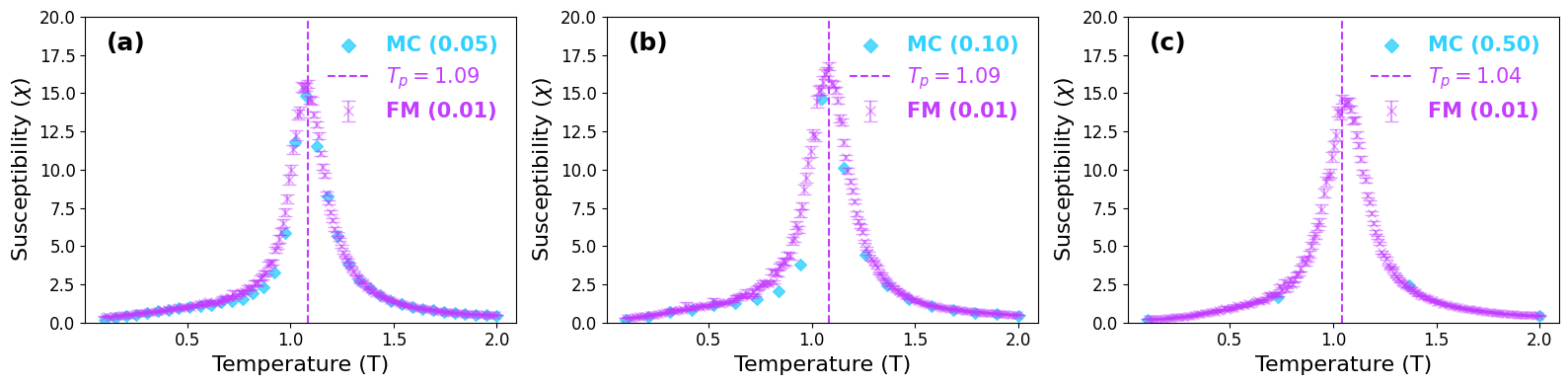}
        \caption{
        Susceptibility $\chi(T)$ for a $32\times32$ XY lattice. The points show results from MCMC (MC) simulations, which serve as both the ground truth and the training data for our Flow Matching (FM) model. The three panels demonstrate the model's performance when trained on this MCMC data at varying levels of precision: (a) a dense grid with temperature spacing $\Delta T = 0.05$, (b) $\Delta T = 0.10$, and (c) a sparse grid with $\Delta T = 0.50$. The training data consists of $5000$ configurations per temperature point. In all cases, the FM model is evaluated on a much finer grid ($\Delta T = 0.01$), showing it captures the qualitative BKT-related structure even when trained on the sparsest data set (in (c), Monte Carlo training data is available only at the four temperatures outside the phase transition point).}
        \label{fig:sus_precision_0_05__0_10__0_50}
    \end{minipage}
\end{figure*}

FM-generated configurations preserve the expected qualitative dependence of physical observables on temperature, even when trained only at a sparse set of temperature points. This diagnostic of initializer quality highlights that we do not use FM as a precision estimator of equilibrium observables on a dense temperature grid. Rather, we ask whether the generated configurations remain sufficiently well aligned with the underlying thermodynamic trends to serve as physically informed warm starts at intermediate temperatures. Fig.~\ref{fig:eng_mag_stiffness_vortex_increase} demonstrates that these FM-generated configurations retain the expected qualitative trends of several observables when evaluated on a denser temperature grid than the one used for training.

This temperature-interpolation property is useful for initialization. Near a transition region, observables such as susceptibility can vary rapidly with temperature, so a practical initializer should remain physically consistent even between sparsely sampled training points. Fig.~\ref{fig:sus_precision_0_05__0_10__0_50} shows that, even when trained on increasingly sparse temperature grids, the FM model still generates configurations whose susceptibility curves track the correct qualitative transition-region structure. We emphasize again that these FM-only observables are not used here to make precision claims about equilibrium critical behavior; rather, they serve as diagnostics to ensure that the generated configurations remain suitable warm starts over a continuous temperature range.

\section{Result II: Cross-Size Trend Consistency of FM-Generated Warm Starts}
\label{sec:results_System_Sizes_Extrapolation}

\begin{figure}[ht]
    \centering
    \includegraphics[width=\textwidth]{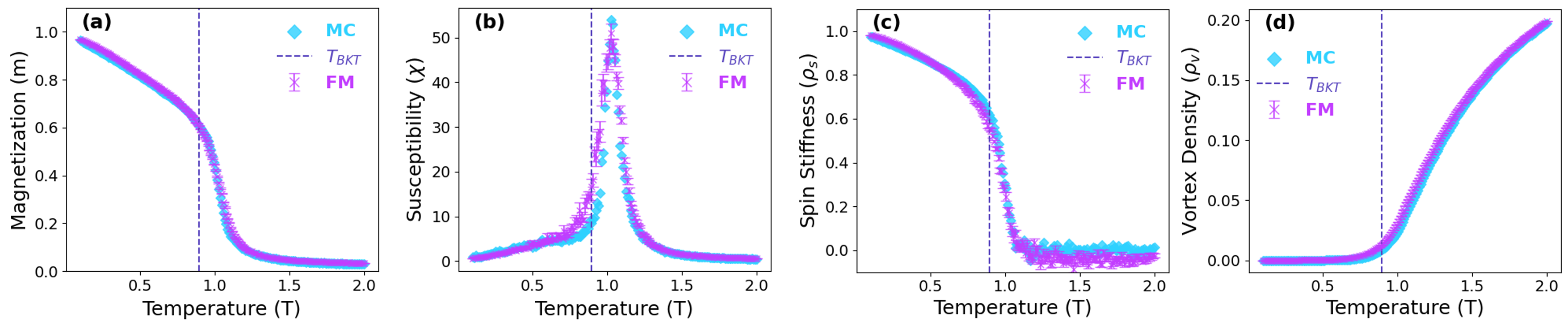}
    \caption{Size extrapolation results for the Flow Matching (FM) model. The model was trained on data from a $32\times32$ system ($T\in[0.1,2.0]$, $\Delta T=0.01$, 1500 configs/T) and then used to generate configurations for a $64\times64$ lattice. The generated observables—(a) magnetization, (b) susceptibility, (c) spin stiffness, and (d) vortex density—are shown to be in qualitative agreement with reference MCMC results for the larger system.}
    \label{fig:size_extrap_eng_mag_sus_stiff}
\end{figure}

A second diagnostic for initializer quality is whether configurations generated on unseen lattice sizes preserve the correct qualitative finite-size trends. This question is central to the warm-start setting considered here: if a model trained on a single lattice size can still produce physically aligned initial configurations on larger lattices, then the same trained initializer can be reused across a broad range of downstream Monte Carlo simulations without retraining. As shown in Fig.~\ref{fig:size_extrap_eng_mag_sus_stiff}, a model trained only on $32\times32$ configurations can be deployed on a $64\times64$ lattice, where the resulting observables remain qualitatively consistent with the reference Monte Carlo trends.

\begin{figure}[ht]
    \centering
    \includegraphics[width=0.5\textwidth]{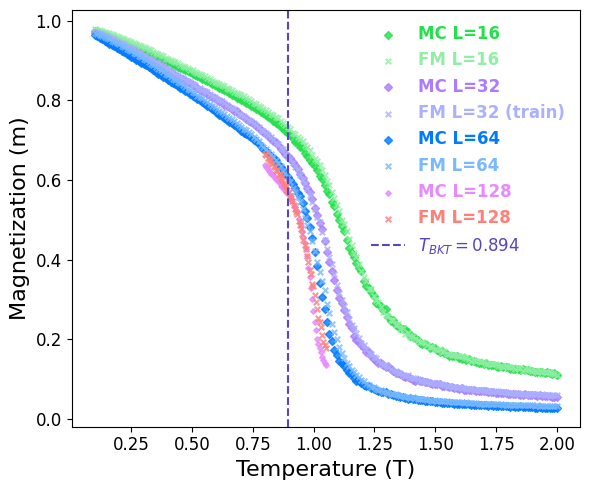}
    \caption{Size extrapolation of magnetization using a single Flow Matching (FM) model trained only on a $32\times32$ lattice ($T\in[0.1,2.0]$, $\Delta T=0.01$, 1500 configs/T). The model generates configurations for various system sizes ($L=16, 64, 128$), and the resulting magnetization is compared against MCMC (MC) data. The FM results match the reference data for each individual size and correctly exhibit the characteristic finite-size scaling trend of the transition.
    } 
    \label{fig:diff_size_mag_results}
\end{figure}

\begin{figure}[ht]
    \centering
    \includegraphics[width=0.5\textwidth]{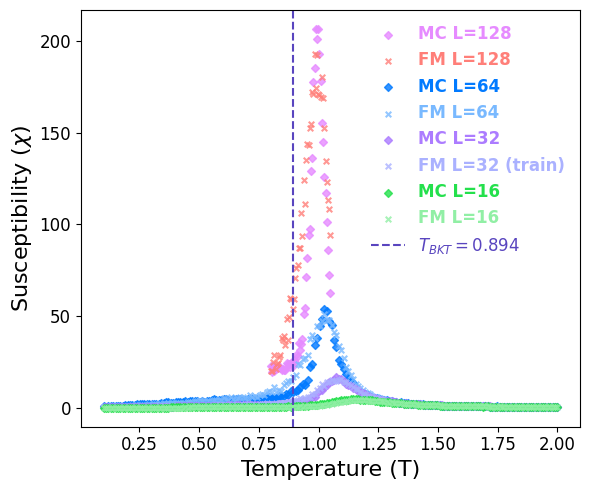}
    \caption{Susceptibility $\chi(T)$ for various system sizes ($L=16, 64, 128$), demonstrating the size-extrapolation capability of a single Flow Matching (FM) model. The model was trained exclusively on $32\times32$ data ($T\in[0.1,2.0]$, $\Delta T=0.01$, 1500 configs/T). Although quantitative deviations from the MCMC (MC) reference increase as the system size diverges from the training size, the FM model qualitatively captures the relevant finite-size trend. Specifically, it correctly reproduces the characteristic shift and sharpening of the susceptibility peak with system size, a key signature of the BKT transition. }
    
    \label{fig:diff_size_sus_results}
\end{figure}

This extrapolation is robust across a wide range of system dimensions. As shown in Figs.~\ref{fig:diff_size_mag_results} and \ref{fig:diff_size_sus_results}, the same model trained on $32\times32$ generates configurations for lattices as small as $16\times16$ and as large as $128\times128$. It is crucial to note that the model was never exposed to the global statistics of the larger systems during training. The emergence of size-dependent features, such as the shifting susceptibility peak, arises because the model applies learned local interaction patterns consistently across the larger lattice geometry. Although quantitative deviations increase with the extrapolation range (likely due to the accumulation of integration errors and the lack of exact long-range constraints in the training data), the model reproduces the essential qualitative finite-size effects and the characteristic drift of BKT-related observables. This interpretation is supported by the explicit receptive-field and correlation-function analysis presented in Appendix~\ref{sec:receptive_field}. For the compact U-Net used here, the theoretical receptive field effectively covers the full $32\times32$ training lattice, but becomes a finite spatial cutoff when the same model is deployed on larger systems. Consistently, the FM-generated spin-spin correlation function remains accurate at short and intermediate distances while visible deviations emerge at longer distances near the transition region. This supports our interpretation that the model transfers reusable local statistical structure across sizes, rather than fully reproducing the long-range critical coherence of the larger equilibrium system.

\begin{figure}[ht]
    \centering
    \includegraphics[width=0.5\textwidth]{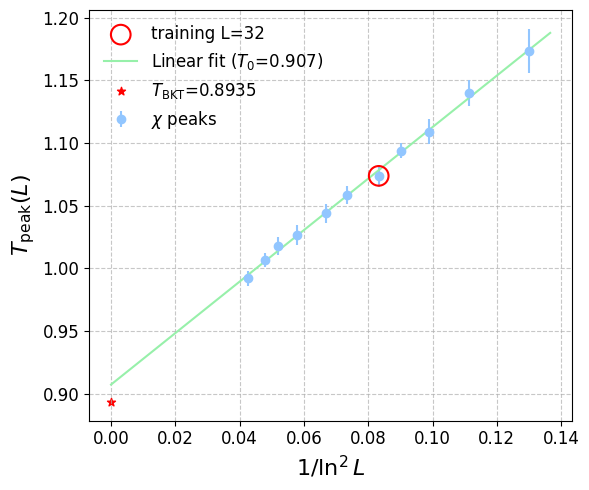}
    \caption{
    Finite-size trend diagnostic based on FM-generated warm-start configurations. 
    The susceptibility peak temperature $T_{\text{peak}}(L)$ is plotted against $(\ln L)^{-2}$ for various lattice sizes ($L=16,\;20,\;24,\;28,\;32,\;40,\;48,\;64,\;80,\;96,\;128$). All points are extracted from configurations generated by a single Flow Matching model trained on $32\times32$ data ($T\in[0.1,2.0]$, $\Delta T=0.01$, 1500 configs/T). {For each lattice size, $T_{\text{peak}}(L)$ was determined from susceptibility curves sampled on a fine temperature grid, and the error bars indicate the statistical uncertainty estimated from 10 independent repetitions with different random seeds.} The approximate linear trend indicates that the generated configurations preserve the expected qualitative finite-size behavior of the transition region. 
    }
    \label{fig:finite_size_scaling}
\end{figure}

We use finite-size scaling here in a deliberately limited sense: not as evidence that FM alone provides a precision route to the thermodynamic-limit BKT transition, but as a stringent trend-consistency test for cross-size initialization. If FM-generated configurations are to be useful warm starts on unseen lattice sizes, they should at least preserve the correct direction of finite-size drift in transition-region diagnostics. This is the role played by the susceptibility peak analysis here.

We note that the spin stiffness and the Nelson--Kosterlitz criterion provide a more rigorous route to precision BKT analysis~\cite{Hsieh_2013_Finite_size_scaling}. However, as discussed in Sec.~\ref {sec:results_error_analysis}, fluctuation-sensitive observables extracted directly from FM-generated ensembles are less reliable, particularly near the transition region and for larger extrapolations in size. For this reason, we do not interpret the present FM-only scaling plot as a definitive estimate of $T_{\mathrm{BKT}}$. Instead, the approximately linear behavior of $T_{\text{peak}}(L)$ versus $(\ln L)^{-2}$ in Fig.~\ref{fig:finite_size_scaling} indicates that the generated warm-start configurations preserve the correct qualitative finite-size trend~\cite{PhysRevB.65.184405}, which is the relevant criterion for the initializer role studied in this work. The extrapolated value $T_{\mathrm{BKT}}\approx0.907$ should therefore be read only as a coarse consistency check, not as a precision claim.

\section{Result III: Error Profile of FM-Only Diagnostics}
\label{sec:results_error_analysis}

\begin{figure}[!ht]
    \centering
    \includegraphics[width=0.5\textwidth]{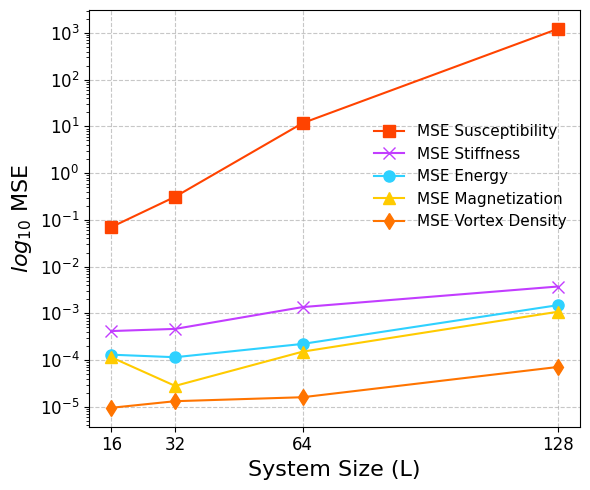}
    \caption{Mean squared error (MSE) as a function of temperature, quantifying the deviation of the size extrapolation method. The MSE is calculated between the Flow Matching and reference MCMC results for several observables: susceptibility $\chi$, spin stiffness $\rho_s$, energy $E$, magnetization $m$, and vortex density $\rho_v$. A single FM model, trained on a $32\times32$ lattice ($T\in[0.1,2.0]$, $\Delta T=0.01$), was used to generate data for systems of size $L=16,\;32,\;64,$ and $128$. The evaluation for $L=16,\;32$ and $64$ uses the same temperature grid as the training, while the evaluation for the largest system, $L=128$, is performed on a finer grid focused on the critical region ($T\in[0.8,1.05]$, $\Delta T=0.005$).
    }
    \label{fig:fm_mc_diff_analysis}
\end{figure}

To quantitatively assess the performance of our Flow Matching model, we compute the mean squared error (MSE) between its generated estimates and the reference MCMC data for key physical observables, as defined in Eq.~\eqref{eq_mse}, where $T$ is temperature and $N_T$ is the number of temperature points. Fig.~\ref{fig:fm_mc_diff_analysis} presents this error analysis for a model trained on $32\times32$ lattices and then used to generate samples for systems of size $L=16,\; 32,\; 64,$ and $128$. Because the FM-generated configurations are not intended to replace equilibrated Monte Carlo samples for precision critical analysis, the error measures in this section should be interpreted as diagnostics of initializer fidelity rather than as the primary performance metric of a standalone simulator.

\begin{align}
    MSE \equiv \frac{1}{N_T} \sum_{T}
    \left[
    \langle\mathcal{O}_{FM}(T)\rangle - \langle\mathcal{O}_{MC}(T)\rangle
    \right]^2
    \label{eq_mse}
\end{align}

The analysis reveals two important trends. First, the deviations for first-moment quantities remain minimal, indicating that the FM-generated warm-start configurations retain sufficient structural information even when extrapolating to system sizes four times larger than the training data. The errors are small for the $L=16$ system, demonstrating robust interpolation to smaller sizes, and predictably increase with the extrapolation range to $L=64$ and $L=128$. Second, and more revealingly, the model exhibits higher consistency for first-moment quantities like energy ($E$) and magnetization ($M$) compared to second-moment quantities related to fluctuations, such as susceptibility ($\chi$) and spin stiffness ($\rho_{s}$). This discrepancy is most pronounced near the critical temperature (see Fig.~\ref{fig:diff_size_sus_results}), where critical fluctuations dominate.

This behavior stems from several factors inherent to the Flow Matching framework. Primarily, as discussed in Sec.~\ref{sec:Flow_of_Random_Sampling}, the model learns to approximate the mean of the conditional velocity field, $\mathbf{u_t}(\mathbf{x}_t) = \mathbb{E}[\mathbf{v_t}(\mathbf{x}_t|\phi)]$. This averaging process inherently smooths the learned flow. And it is a direct consequence of the bias-variance tradeoff introduced by the mean squared loss function used for training. According to the loss function in Eq.~\eqref{eq_loss_fn_1}, the expected error of the learned velocity field $\mathbf{v_\theta}$ can be decomposed as:

\begin{align}
    \mathbb{E}[(\mathbf{v_\theta}-\mathbf{v_t})^2] 
    = \left(\mathbb{E}[\mathbf{v_\theta}-\mathbf{v_t}]\right)^2 
    + \text{Var}[\mathbf{v_\theta}-\mathbf{v_t}]
\end{align}

\noindent To minimize this total error, the model must balance its variance (due to finite training data) against a systematic bias towards the mean of the true conditional velocity field. This optimization naturally produces a learned vector field that prioritizes the mean field trend over the precise variance of the target distribution. Furthermore, this initial inaccuracy in the learned field's variance is compounded during the inference stage. Sampling requires numerical integration of the probability flow ODE with a finite number of steps. As suggested in Fig.~\ref{fig:naive_flow_matching}, the vector guiding the trajectory in each discrete step of the numerical solver is an approximation. Minor errors in the velocity field, particularly regarding its variance, accumulate during the iterative integration process. This can cause the final FM-generated warm-start configurations to underestimate the actual spread of the target distribution, leading to a larger error in the variance of the resulting ensemble. Consequently, while the generated samples accurately reflect first-moment quantities such as energy and magnetization, the fluctuation-related properties of the generated ensemble are less reliable. A complementary, architecture-level explanation is provided by the receptive-field analysis in Appendix~\ref{sec:receptive_field}. There, we show that, on extrapolated lattices, the compact U-Net accurately reproduces spin-spin correlations mainly within the spatial scale accessible to its receptive field, while longer-distance correlations near the transition region are progressively underestimated. This provides a mechanistic explanation for why local or first-moment observables remain comparatively robust, whereas fluctuation-sensitive quantities that depend more strongly on system-spanning coherence, especially spin stiffness, show larger deviations. Despite FM-generated samples showing larger errors in fluctuation-related physical statistics, which grow more significantly with size extrapolation, the model still preserves the qualitative location of characteristic susceptibility peaks across system sizes. This makes the model useful as a physically informed initializer whose generated configurations already lie in the correct qualitative transition region, even though the absolute magnitude of fluctuation-sensitive observables is not sufficiently accurate for standalone precision analysis.

\begin{figure}[!ht]
    \centering
    \includegraphics[width=0.6\textwidth]{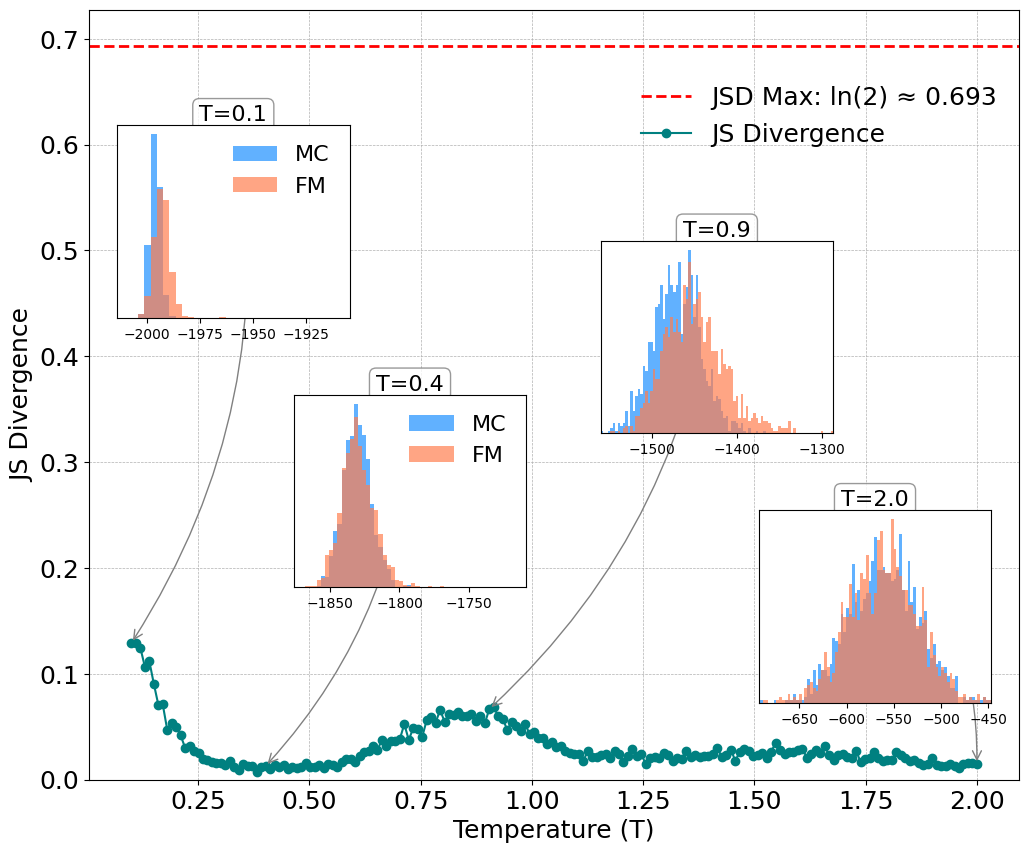}
    \caption{Jensen–Shannon Divergence (JSD) between the energy distributions generated by the Flow Matching (FM) and MCMC methods as a function of temperature. The analysis was performed on a $32\times32$ lattice, using an FM model trained and evaluated under the specified conditions ($T\in[0.1,2.0]$, $\Delta T=0.01$, 1500 configs/T). The insets show the corresponding energy histograms at four representative temperatures. All histograms are plotted with identical bin widths to compare the distributions' spread visually. As can be seen in the figure, JSD has a significant value near zero temperature ($T = 0.1$) and the phase transition point ($T\approx 0.89$), which means that the neural network still cannot perform perfectly in the extreme distribution region, but over most of the temperature range it remains sufficiently close to the Monte Carlo reference to support the use of FM-generated samples as physically informed warm-start configurations for subsequent Monte Carlo refinement.
    }
    \label{fig:js_divergence}
\end{figure}

In the training size ($32\times 32$), we also analyze the distribution error of the energy versus the different temperature. We use Jensen--Shannon divergence (JSD) as a measure of the difference between Flow Matching and MCMC sampling distributions. For two discrete probability distributions $\mathcal P$ and $\mathcal Q$ in the same domain $\mathcal X$ define the midpoint distribution.

\begin{align}
\mathcal M \equiv \tfrac12\bigl(\mathcal P + \mathcal Q\bigr).\label{eq:Mmid}
\end{align}
The Jensen--Shannon divergence (JSD) is given by
\begin{align}
\operatorname{JSD}(\mathcal P\|\mathcal Q)
\equiv 
\tfrac12 \sum_{x\in\mathcal X} \mathcal P(x)\log\frac{\mathcal P(x)}{\mathcal M(x)}
+
\tfrac12 \sum_{x\in\mathcal X} \mathcal Q(x)\log\frac{\mathcal Q(x)}{\mathcal M(x)}.\label{eq:JSD}
\end{align}
With natural logarithms, $0\le\operatorname{JSD}(\mathcal P\|\mathcal Q)\le\ln 2$, the upper bound $\ln 2$ is reached precisely when $\mathcal P$ and $\mathcal Q$ are completely unrelated. Fig.~\ref{fig:js_divergence} shows the JSD of energy distribution at different temperatures. Across most of the temperature range, including the phase transition point ($T\approx 1.0$ for system size $32 \times 32$), the JSD is less than 0.1, indicating that the generated energy distributions are statistically consistent with the Monte Carlo reference. Only near zero temperature does the JSD exceed 0.1. The histograms embedded in Fig.~\ref{fig:js_divergence} show that the energy distribution becomes increasingly concentrated as the temperature decreases (the bar widths in all histograms are the same). Because Flow Matching uses only one neural network model that is simultaneously trained and sampled over the entire temperature range, the figure reveals that under the same conditions, the generative model faces challenges in resolving the fine-grained variance of extremely narrow distributions. The bias of distributions near the phase transition point also indicates that the neural network is still some distance from generating distributions that are entirely consistent with physics. Nevertheless, the generated distributions remain sufficiently close to the Monte Carlo reference over most of the temperature range to support the use of FM-generated samples as physically informed warm starts for subsequent Monte Carlo refinement.

\section{Result IV: Amortized Computational Cost of Reusable FM Initialization}
\label{sec:results_Time_Cost_Comparison}
The computational motivation for the present framework is the reuse of a single trained FM model as an initializer across many temperatures and lattice sizes. In this section, we therefore compare the amortized cost of reusable FM generation against the cost of running separate Monte Carlo simulations from scratch on each lattice size. This comparison does not, by itself, constitute an end-to-end benchmark of hybrid FM--MCMC equilibration speedup; rather, it isolates the potential computational advantage of replacing uninformed initialization with reusable, physically informed warm-start generation. For our FM model, this involves a one-time training cost on a small lattice ($32\times 32$), followed by rapid inference on a series of larger lattices. The MCMC method requires running a separate, full simulation for each lattice size. We use our highly optimized, GPU-accelerated MCMC algorithm (see Appendix~\ref{app:mc_gpu}) as the baseline for this comparison. All timing benchmarks were performed on a single workstation; detailed hardware and software specifications are provided in Appendix~\ref{app:computational_details}. The key metric for this evaluation is the Sampling Cost, defined as the average wall-clock time to generate a single independent configuration:

\begin{align}
    \text{Sampling Cost} \equiv \frac{t_{\text{sample}}}{N_{\text{sample}}}
    \label{eq_time_cost_sampling}
\end{align}
where $t_{\text{sample}}$ is the total time required to generate $N_{\text{sample}}$ independent samples across all temperatures and ensembles.

For the FM model, the overall cost has two components: the inference cost, given by Eq.~\eqref{eq_time_cost_sampling}, and a one-time, upfront training cost. The average time per configuration during the training phase is defined as:
\begin{align}
    \text{Training Cost} \equiv \frac{t_{\text{train}}}{N_{\text{train}}}
    \label{eq_time_cost_training}
\end{align}
where $t_{\text{train}}$ is the total time to process $N_{\text{train}}$ training configurations. The training cost here is the amortized time over the total training samples. 

\begin{figure}[!ht]
    \centering
    \includegraphics[width=0.5\textwidth]{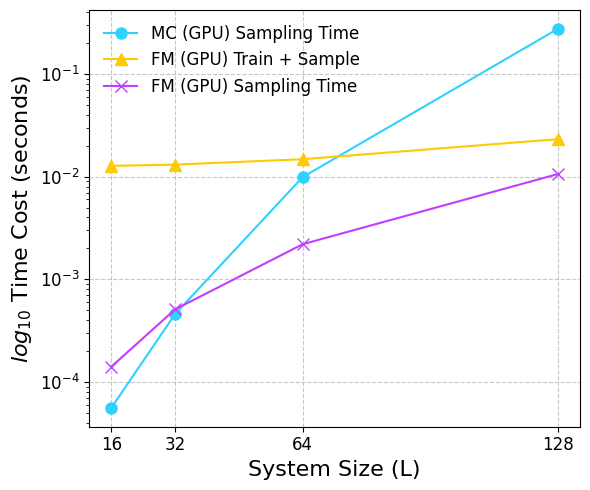}
    \caption{Computational cost per independent configuration as a function of system size $L$. The figure compares the performance of our GPU-accelerated MCMC method against the Flow Matching (FM) model, which was trained only on the $L=32$ system. The FM cost is presented in two ways: the inference cost alone and the total cost including the amortized one-time training expense, as defined in Eq.~\eqref{eq_time_cost_training}. The results show a crossover in efficiency, which can be used to generate reusable warm-start configurations across multiple lattice sizes. }
    \label{fig:time_cost_comparison}
\end{figure}

As shown in Fig.~\ref{fig:time_cost_comparison}, while the initial training constitutes a significant upfront investment, this cost is amortized over all subsequent sample generation for various larger system sizes. The crossover point between $L=64$ and $L=128$ indicates that the one-time training cost can be amortized once the trained model is reused across sufficiently many larger systems. From the initializer perspective adopted in this work, the key implication is not that FM replaces equilibrium Monte Carlo, but that it can shift part of the initialization cost into an offline, reusable stage. This highlights that the actual efficiency of our model lies not in single-sample speed, but in its ability to circumvent the need for repeated, costly simulations across different scales.

\section{Why FM Warm Starts Transfer Across System Sizes}
\label{sec:disscuss_System_Size_Extrapolation}

The cross-size transfer observed in this work should be interpreted cautiously. Our claim is not that the model exactly reproduces the full equilibrium long-range structure of unseen larger systems. Rather, we argue that the combination of the FM objective and the U-Net architecture allows the model to learn local statistical regularities that are sufficiently reusable across lattice sizes to generate physically aligned warm-start configurations. The following subsections explain why such a transfer is plausible from the perspective of locality, translation equivariance, and multi-scale feature processing.

\subsection{Inductive Bias}
Inductive bias refers to implicit assumptions within machine learning algorithms that enable generalization to unseen data. The effectiveness of an algorithm depends significantly on how well its inductive biases align with the characteristics of the given task. Thus, selecting an appropriate model involves matching these biases to the specific nature of the problem. For example, several recent studies similar to this work have pointed out that the equivariance of neural networks (a kind of inductive bias) has significant advantages in dealing with physical systems with symmetries~\cite{Schuh:2025fky_Hubbard_Model_NF, Equivariant_flow_matching}.

Convolutional Neural Networks (CNNs) inherently contain strong inductive biases that are advantageous for processing natural images whose statistical properties align closely with these biases. Consequently, CNNs efficiently learn visual representations with minimal data. These biases are equally effective in physical systems characterized by local interactions.
The core computational mechanism of CNNs is convolution, typically implemented using small convolutional kernels (e.g. $3\times3$ in this work). This design explicitly encodes the inductive bias of locality, which posits that the most relevant information for interpreting any given part of an image is predominantly contained within its spatial vicinity. Thus, the significance of any pixel largely depends on its neighboring pixels, allowing CNN neurons to focus solely on small, localized receptive fields. This notion parallels the physical treatment of complex many-body systems, where interactions are often defined locally. Physicists commonly represent such systems using local Hamiltonians, and physical fields or spin configurations governed by these Hamiltonians naturally conform to CNN's locality inductive bias, where the state or behavior at a given location is determined by its neighbors.

A second crucial inductive bias inherent in CNNs is translation equivariance, arising from parameter sharing across convolutional layers. Within each convolutional layer, the same kernel systematically traverses the entire input feature map, ensuring that the features and behaviors learned at any specific location can directly inform the interpretations at all other spatial locations. This bias aligns with the practice in physics of using uniform or similarly structured local Hamiltonians to describe complex many-body systems consistently throughout their spatial domain, rather than employing entirely distinct physical mechanisms or Hamiltonians at different locations.

The powerful inductive biases embedded in CNN architectures, particularly locality and translation equivariance, are pivotal to their efficiency in image processing tasks. They also intrinsically resonate with the fundamental properties and representations of physical many-body systems when dealing with the data at the physical field or spin configuration level. 

\subsection{U-Net Structure}
\begin{figure}[!ht]
    \centering
    \includegraphics[width=0.5\textwidth]{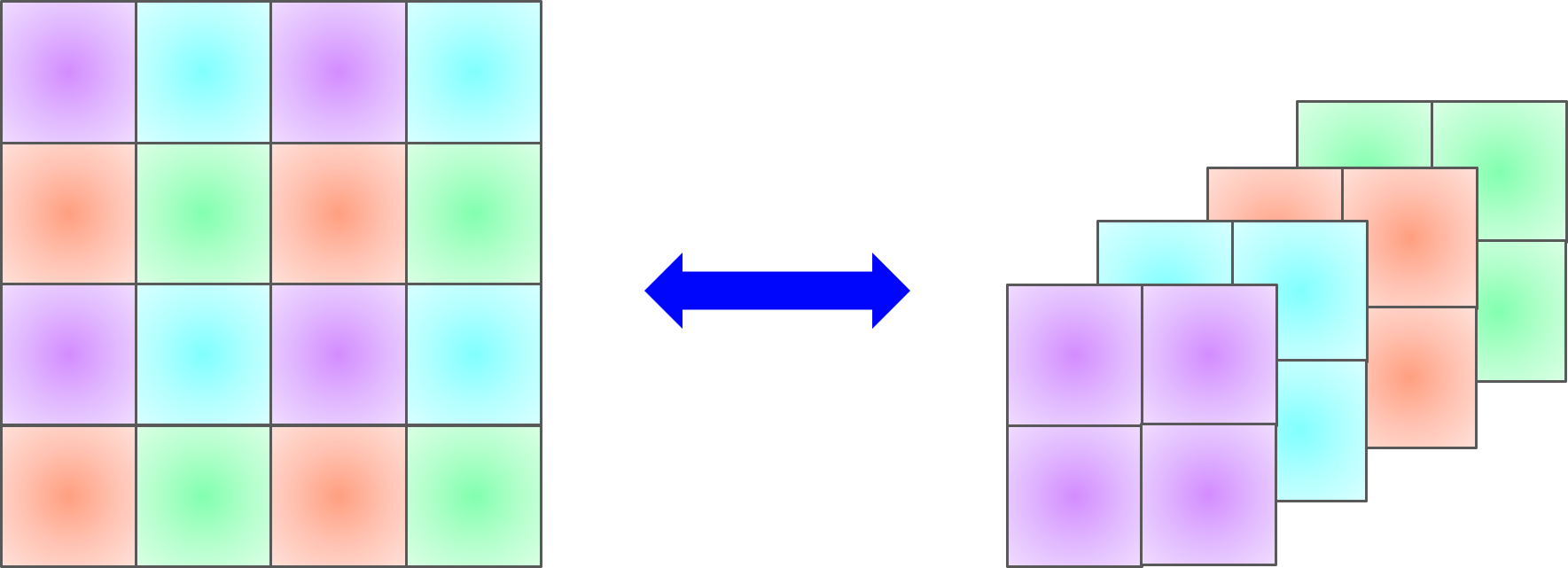}
    \caption{The pixel shuffle operation~\cite{shi2016pixel_shuffle}, a lossless resizing technique used in our U-Net architecture. The encoder employs a space-to-depth pixel shuffle (illustrated from left to right) to downsample spatial resolution while increasing channel depth. Conversely, the decoder uses an inverse depth-to-space pixel shuffle (right to left) for upsampling to reconstruct the spatial dimensions.}
    \label{fig:pixel_shuffle}
\end{figure}

While the inductive biases of locality and translation equivariance make standard CNNs well suited to systems with local interactions, they struggle with long-range correlations near criticality because the receptive field grows only linearly with depth of network. The U-Net architecture~\cite{ronneberger2015unet}, built by the encoder and decoder structure, mitigates this by efficiently integrating information across multiple spatial scales, especially when paired with information-preserving resizing (e.g., pixel-shuffle–based rearrangements (see Fig.~\ref{fig:pixel_shuffle})~\cite{shi2016pixel_shuffle}). Crucially, even when operating on downsampled feature maps in the encoder that summarize large-scale structure, convolution remains a local operator: kernels still act on neighboring sites of the current lattice, thereby learning transformation rules that are consistent across resolutions—i.e., effective local interactions in renormalized coordinates that remain (approximately) scale-invariant. To ensure that the model accurately reflects the underlying physics, we align its architectural choices with the simulation's boundary conditions. All Monte Carlo simulations in this work employ periodic boundary conditions. Correspondingly, all convolutional layers within the U-Net architecture are implemented with circular padding. This crucial detail enforces the same periodic topology on the feature maps as on the physical lattice, ensuring that the learned local rules are consistent with the system's global structure.

Mechanistically, the encoder uses an pixel shuffle (space-to-depth)~\cite{shi2016pixel_shuffle} to reduce spatial resolution without discarding information, mapping, for example, each $2\times2$ block into four channels (see Fig.~\ref{fig:pixel_shuffle}). This preserves translation equivariance and expands the effective receptive field exponentially with depth~\cite{wang2024theoreticalanalysis_inductive_biases}, while convolutions at coarser resolutions still implement local updates on block variables. The decoder applies the inverse pixel shuffle (depth-to-space) to reconstruct high-resolution outputs (see Fig.~\ref{fig:pixel_shuffle}), and skip connections concatenate encoder features with aligned decoder features. This fusion preserves fine-grained locality while enforcing cross-scale consistency, allowing the network to integrate global context without sacrificing the learning of scale-consistent local rules at each resolution. For more architectural and technical details, see the Appendix~\ref{app:nn}.

For many-body physical systems, this design naturally aligns with a renormalization-group perspective: space-to-depth provides a lossless coarse-graining that exposes emergent long-range correlations at the bottleneck, while convolutions at every level learn local interaction laws on the corresponding (reindexed) lattice. Skip connections then reinsert microscopic details, ensuring that global, system-spanning patterns inform but do not overwrite local physics. As a result, the U-Net can combine local information with broader contextual cues, which is sufficient in our experiments to preserve qualitative cross-size trends in the generated warm starts. This locality-dominated interpretation is not merely heuristic. In Appendix~\ref{sec:receptive_field}, we compute the theoretical receptive field of the present U-Net and compare it directly with spin-spin correlation functions on both the training lattice and an extrapolated larger lattice. The analysis shows that the model behaves nearly as a global operator on the $32\times32$ training system, but imposes a finite spatial cutoff when deployed on larger lattices. This explains why the network can preserve qualitative finite-size trends while still failing to reproduce the full long-range phase coherence required for precision critical observables.

To empirically validate that the U-Net's inductive biases are key to its successful size extrapolation, we conducted a comparative study against a Diffusion Transformer (DiT) model~\cite{peebles2023DiT}, which lacks strong, built-in locality. The U-Net significantly outperformed the DiT, especially in the extrapolation task, despite having substantially fewer parameters. This result supports our claim that architectural alignment with the problem's local nature is critical. The detailed experimental setup, results, and discussion of this comparison are presented in Appendix~\ref{app:unet_vs_dit_comparison}.

\subsection{Capturing Local Dynamics for Scalable Generation}
In the initializer setting considered here, the relevant distinction is not that FM should replace MCMC, but that the two methods play different roles. MCMC remains the equilibrium engine that produces rigorously correct Boltzmann samples after sufficient equilibration. Flow Matching, by contrast, is used here as a reusable generator of warm-start configurations. Its task is therefore much weaker but also more computationally attractive: it need not reproduce the full equilibrium ensemble with high precision on its own, but only generate states that already lie close enough to the relevant physical region that subsequent MCMC equilibration becomes easier.

From this perspective, the usefulness of FM across system sizes comes from learning local statistical regularities rather than memorizing size-specific global ensembles. For a model such as the XY system, whose Hamiltonian is local, many of the short-range structural cues that distinguish low-temperature, transition-region, and high-temperature configurations are also local. A convolutional architecture can therefore reuse these learned local rules on lattices of different sizes, producing configurations that remain qualitatively well aligned with the correct finite-size trends.

The principle of FM is to construct a continuous path that transforms a simple, unstructured probability distribution (e.g., Gaussian noise) into the complex target data distribution (in our case, the physical spin configurations from MCMC). The characteristics of this generative framework include:

\begin{enumerate}
    \item \textbf{Directed Dynamics :}
    The process is governed by an ODE defined by a learned vector field. This vector field acts as a time-dependent force that deterministically "flows" any initial random state $\mathbf{x}_0$ at time $t=0$ to a structured physical state $\mathbf{x}_1$ that is statistically consistent with the physical ensemble at time $t=1$. Unlike MCMC, this process is explicitly designed to change the probability distribution over time, moving it along a predefined path from noise to data.
    
    \item \textbf{Approximating Local Dynamics of Transformation :} 
    The model is trained to minimize the discrepancy between its vector field and a target vector field derived from the MCMC data. Crucially, the model is not learning the global statistical properties of the small training systems. Instead, it is learning a function that, for any given configuration $\mathbf{x}_t$ at any intermediate time $t$, dictates the optimal infinitesimal change required to align the configuration with the target distribution. For the XY model, this involves learning local update tendencies, such as encouraging adjacent spins to align to lower the energy or resolving topological defects, such as vortices.

    \item \textbf{Distillation of Scale-Invariant Patterns:} 
    The Hamiltonian of the XY model is defined by local, nearest-neighbor interactions. The physical principles governing low-energy configurations, the preference for smooth spin variations, and the structure of vortices are therefore local and independent of the overall system size. Flow Matching, by learning the vector field from local configuration data, distills these scale-invariant interaction patterns. Powered by the inductive biases embedded in CNN architectures, the learned vector field at a specific site effectively only "sees" the configuration of its local neighborhood, not the global lattice boundaries.
\end{enumerate}

This viewpoint also clarifies how qualitative finite-size effects can appear without implying exact thermodynamic-limit fidelity. The model is not assumed to encode the full long-range equilibrium structure of each unseen lattice size. Rather, when the same learned local statistical rules are deployed on larger periodic lattices, the generated configurations can still display the correct direction of finite-size drift in coarse observables. This is sufficient for the warm-start role studied in the present work.

Accordingly, we interpret the success of cross-size deployment as evidence that the model has learned reusable local statistical structure for initialization, not as proof that it has become a standalone simulator of critical many-body physics. In our view, this is precisely the practically relevant regime for physics-informed machine learning in this context: learning enough of the local structure to initialize downstream exact methods efficiently. This success underscores a powerful principle in physics-informed machine learning: the most effective models for physical problems are often those whose architectural inductive biases are deliberately aligned with the intrinsic locality and symmetries of the physical system being studied, a principle that has proven effective in a growing body of literature at the intersection of machine learning and physics~\cite{Gerdes_2023LQF_equi_flow, Schuh:2025fky_Hubbard_Model_NF, Equivariant_flow_matching, PhysRevLett.125.121601, PhysRevD.103.074504, Hashimoto_2024}.

\section{Discussion}
\label{sec:discussion}
\subsection{Comparison of Normalizing Flows and Flow Matching}
\label{sec:comparison}
\begin{figure}[!ht]
    \centering
    \includegraphics[width=0.5\textwidth]{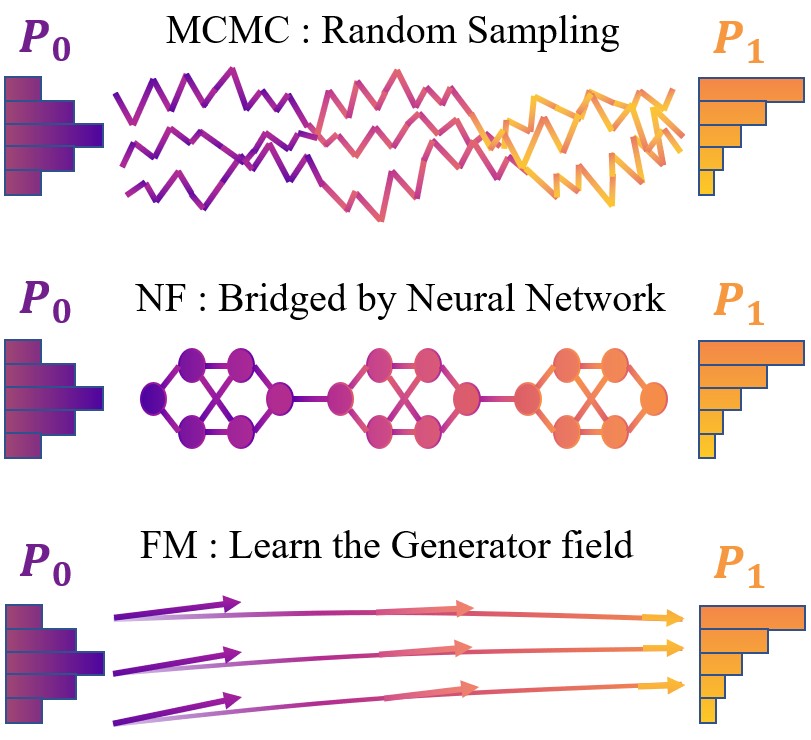}
    \caption{A schematic comparison of three sampling methodologies. Traditional Markov Chain Monte Carlo (MCMC) operates by randomly sampling configurations from a system's phase space according to probability weights, a process that can be computationally intensive. Normalizing Flows (NFs) offer a generative alternative by constructing a sequence of invertible maps, but their reliance on a tractable Jacobian determinant imposes significant constraints on the neural network architecture. In contrast, Flow Matching (FM), the approach used in this work, directly learns the vector field (flow) that generates the transformation between distributions. This allows for greater architectural flexibility and a more stable training objective, as samples are transported deterministically along a predefined path.} 
    \label{fig:mcmc_nf_fm}
\end{figure}

Both normalizing flows (NFs) and Flow Matching (FM) are powerful generative frameworks designed to learn a transformation between a simple, tractable base distribution and a complex target data distribution. However, they differ fundamentally in their approach, leading to significant implications for model architecture, training stability, and flexibility (see Fig.~\ref{fig:mcmc_nf_fm}).

Normalizing Flows construct this transformation as a sequence of invertible and differentiable mappings. This framework's core requirement is that each transformation's Jacobian determinant must be computationally tractable. This is essential because NFs rely on the change-of-variables formula to compute the exact likelihood of the target distribution, which is typically used to minimize the Kullback-Leibler (KL) divergence during training. This constraint imposes severe restrictions on the neural network architectures that can be employed. To ensure a tractable Jacobian, NFs often utilize specialized structures like coupling layers~\cite{real_nvp, nice}, which partition the input dimensions into active and passive sets. Consequently, these models require carefully designed problem-specific masking mechanisms to ensure that all dimensions are updated over a sequence of layers. This architectural rigidity makes the implementation more complex and potentially unstable, and may also break the inherent symmetries of the modeled system~\cite{PhysRevLett.125.121601, PhysRevD.103.074504}, a significant drawback in physical applications.

Flow Matching, on the other hand, circumvents these limitations by reframing the problem. Instead of learning a series of discrete, invertible maps, FM learns the continuous-time vector field—the generator—of a probability flow that transports the base distribution to the target distribution. The model is trained via a straightforward regression objective: minimizing the mean squared error between the network's predicted vector field and a target vector field derived from simple interpolations between noise and data. Since this process does not require the computation of Jacobian determinants during training, it is free from the rigid architectural constraints that limit NFs. This allows highly flexible and powerful architectures, such as the U-Net employed in this work, without the need for intricate masking schemes or concerns about invertibility. The resulting training process is typically more stable and scalable, highlighting a key advantage of FM in building robust and generalizable generative models. By focusing on learning the flow generator, FM provides a more direct and versatile framework, which we will discuss further.

\subsection{Generator Learning Framework}
\label{sec:Generator_Learning_Framework}
For the purposes of this work, the relevance of the generator-learning viewpoint is not that it replaces equilibrium simulation, but that it enables a reusable mapping from noise to physically informed warm-start configurations.
Learning the generator in Flow Matching methods, specifically the velocity field along the probability path connecting different distributions, provides numerous advantages. By explicitly modeling this generator through a neural network, the mapping between distributions becomes explicit and mathematically tractable, circumventing the implicit representations typically encountered in conventional neural networks. This explicit representation significantly improves the flexibility and applicability of the neural network model. Furthermore, framing the neural network as the generator of the distribution mapping, it becomes feasible to employ diverse training strategies beyond Flow Matching alone, such as traditional maximum likelihood estimation or reverse KL divergence minimization in training Normalizing Flows (but need to solve the probability flow ODE during the training process), which can also be adapted to suit various problem contexts and specific research objectives. Overall, this approach aligns well with contemporary efforts in machine learning to build more interpretable, flexible, and versatile generative models, enabling clearer insight into the learned representations and improving the model's adaptability across a broader range of applications.

\subsection{Limitations of Flow Matching}
 \label{sec:limitation} 

 A significant limitation highlighted by our analysis is the model's tendency to underestimate second-moment observables, which are directly related to physical fluctuations. This is evident in the increased MSE for susceptibility as system size grows (see Fig.~\ref{fig:fm_mc_diff_analysis}) and, more strikingly, in the statistical noise observed in the extrapolated spin stiffness, as detailed in Appendix~\ref{app:more_observables}. This behavior is likely rooted in the nature of the regression-based training objective (Eq.~\ref{eq_loss_fn_1}), which prioritizes the mean-field trend of the conditional velocity field, often at the expense of capturing the full variance. 

An additional limitation comes from the finite spatial scale accessible to the compact U-Net architecture used in this work. As shown by the receptive-field and correlation-function analysis in Appendix~\ref{sec:receptive_field}, the model reproduces local and intermediate-range correlations well, but progressively loses fidelity for longer-range correlations when deployed on lattices larger than the training size. This architectural constraint is consistent with the observed degradation of fluctuation-sensitive observables, especially those that depend on macroscopic phase coherence. This characteristic limits the utility of FM-generated ensembles for standalone precision analysis of fluctuation-sensitive observables.

However, as discussed in Sec.~\ref{sec:results_System_Sizes_Extrapolation}, the FM-generated configurations still preserve the qualitative finite-size trends needed for their use as warm-start states. Improving the model's ability to reproduce the variance structure of the target distribution more faithfully is therefore an important direction for future work. While potential avenues could include exploring alternative loss functions, a more immediate and practical direction is to integrate the method into a hybrid FM--MCMC workflow. In this paradigm, the configurations generated by the FM model serve as physically informed initial states for subsequent MCMC refinement. Subsequent MCMC refinement can be used to correct residual biases from the generative stage, while the final equilibrium accuracy remains the responsibility of the Monte Carlo simulation itself. In this workflow, FM is used to provide reusable warm-start configurations across temperatures and lattice sizes, while MCMC remains the method that restores equilibrium statistics when precision observables are required. From this perspective, the FM stage functions as a reusable initializer that can amortize part of the cost of preparing simulations across temperatures and lattice sizes.

Another limitation is that the Flow Matching framework is designed for systems with continuous degrees of freedom, as in the XY model. Its extension to discrete models (e.g., Ising or Potts) is non-trivial, as the continuous probability flow is not directly applicable. There are two main paths to address this challenge. The first involves reformulating the discrete problem into an equivalent continuous one, for instance, using an auxiliary field method like the Hubbard-Stratonovich transformation, making the system tractable for our approach. A second, more direct path is to leverage generative models explicitly designed for discrete spaces. The development of frameworks such as Masked Diffusion~\cite{masked_diffusion} and more recent Discrete Flow Matching~\cite{gat2024discrete_flow_matching} marks a promising research frontier for applying these powerful sampling techniques to a broader class of physical systems. Indeed, several recent works have already started to explore such applications~\cite{tuo2025scalablemultitemperaturefreeenergy}.

\section{Conclusion and Outlook}
\label{sec:conclusion_and_outlook}

In this work, we have repositioned Flow Matching in a deliberately modest but practically useful role: not as a standalone replacement for equilibrium Monte Carlo, but as a scalable \emph{physically informed initializer} for many-body simulations. Using the 2D XY model as a benchmark, we showed that a U-Net-based FM model trained on small-system data can be reused across temperatures and deployed on larger lattices to generate warm-start configurations that preserve the correct qualitative physical trends.

Under this interpretation, observables computed directly from FM-generated samples are not the final scientific output of the method. Instead, they serve as diagnostics of initializer quality, indicating whether the generated configurations are sufficiently well aligned with the relevant physical regime to support downstream Monte Carlo refinement. Our results show that this criterion is met at the qualitative level across both temperature interpolation and cross-size transfer, even though fluctuation-sensitive observables such as susceptibility and spin stiffness remain insufficiently accurate for standalone precision critical analysis.

The main value of the present framework is therefore computational reuse. A single trained FM model can be amortized across multiple temperatures and lattice sizes, shifting part of the initialization burden from repeated online Monte Carlo equilibration into a reusable offline learning stage. In future work, the most important next step is to quantify the end-to-end gain of this hybrid FM--MCMC workflow directly, for example, through burn-in, equilibration time, and effective sample size benchmarks. Improving variance fidelity in the FM stage---through alternative objectives, variance-aware training, or physics-informed constraints---is another promising direction, but such improvements should be viewed as enhancing a warm-start accelerator rather than redefining the role of exact Monte Carlo in precision many-body simulation. Viewed in this way, the value of FM lies in making transition-region studies more scalable through reusable, physically informed initialization.

%
%

\ack{We thank Chi-Ting Ho, Hsuan-Yu Wu, and Po-Chung Chen, for valuable discussions and suggestions. During the preparation of this work, the authors used ChatGPT and Google Gemini to improve the language and readability of the manuscript. After using this tool, the authors reviewed and edited the content as needed and take full responsibility for the content of the published article. This tool was not used to create, modify, or manipulate original research data or results.}

\funding{This work is supported by the National Center for Theoretical Sciences, the Higher Education Sprout Project funded by the Ministry of Science and Technology, and the Ministry of Education in Taiwan. DWW is supported under the grant NSTC 113-2112-M-007-036.}


\data{The source code for this work's models, simulations, and analysis is publicly available. The code can be accessed from the GitHub repository at \href{https://github.com/ToelUl/Flow-to-Field}{https://github.com/ToelUl/Flow-to-Field}.}


\appendix

\section{Details of Loss Function}
\label{app:details_of_loss_fn}

We can show that the gradient of the conditional loss in Eq.~\eqref{eq_loss_fn_1} is identical to that of the original objective in Eq.~\eqref{eq_loss_fn_0}:

\begin{align}
    \nabla_{\theta}\mathcal{L}_{FM} &= 
    \int_{0}^{1}dt \int \left[\mathcal{D}\mathbf{x}_t\right] p_t(\mathbf{x}_t)
    \nabla_{\theta}\left[\mathbf{v_{\theta}}^2 - 2\mathbf{v_{\theta}}\cdot\mathbf{u_t}\right] \nonumber\\
    &=\int_{0}^{1}dt \int \left[\mathcal{D}\mathbf{x}_0\mathcal{D}\phi\right]
    \nabla_{\theta}\left[
    p_t(\mathbf{x}_t|\phi)q(\phi)\mathbf{v_{\theta}}^2 \right. 
    \left.+\;2\mathbf{v_{\theta}}\cdot\mathbf{v_t}(\mathbf{x}_t|\phi)p_t(\mathbf{x}_t|\phi)\; q(\phi)
    \right] 
    =\nabla_{\theta}\mathcal{L}_{CFM}
\end{align}

\noindent However, it is numerically difficult to calculate the loss function $\mathcal{L}(\theta)$ in Eq.~\eqref{eq_loss_fn_1} directly. Thus, the strategy is to sample few pairs of $(\mathbf{x}_0, \mathbf{x}_1)$ and other conditions randomly in each training step, calculate the corresponding $\mathbf{x}_t$, and minimize $\left[\mathbf{v_{\theta}}(\mathbf{x}_t,t|\phi) - \mathbf{v_t}(\mathbf{x}_t|\phi)\right]^2$ by updating the parameters $\theta$. After repeating the above steps until the neural network converges, we can use learned neural network $\mathbf{v_{\theta}}(\mathbf{x}_t,t|\phi)$ to approximate velocity field $\mathbf{u_t}(\mathbf{x}_t)$. Now the loss function is the expectation value that can be written as 

\begin{align}
    \mathcal{L}_{CFM}(\theta) =  
    \mathbb{E}_{(\mathbf{x}_1, \phi) \sim \mathcal{P}_{data},\; \mathbf{x}_0 \sim \mathcal{N}(0, I),\;t\sim p(t)}
    \left[\mathbf{v_{\theta}}(\mathbf{x}_t,t|\phi) - (\mathbf{x}_1-\mathbf{x}_0)\right]^2
    \label{eq_loss_fn_fm}
\end{align}

\noindent Here, the distribution of time $p(t)$ can be chosen as uniform between 0 and 1. But the better choice of $p(t)$ will be discussed in the next section. 

\section{Training and Sampling Strategy}
\label{app:train_and_sampling_strategy}

According to the path defined in Eq.~\eqref{eq_interpolation_path}, we can define the effective signal-to-noise ratio (SNR) and the logarithm SNR ($\lambda(t)$) as 

\begin{align}
    SNR \equiv \frac{t^2}{(1-t)^2} \;\;,\;\; \lambda(t) \equiv\log SNR = 2 \mathrm{logit}(t)  
    \label{eq_snr}
\end{align}

\noindent Experience has revealed that Flow Matching will have a relatively large error when the SNR is 1, so sampling time from a logit normal distribution in Eq.~\eqref{eq_logit_normal} with $\mu=0$ and $\sigma=1$ during the training phase would be a good choice.\cite{esser2024SD3}

\begin{align}
    p(t, \mu, \sigma) \equiv
    \frac{1}{\sigma \sqrt{2\pi}}\frac{1}{t(1-t)}
    \exp\left(-\frac{\left(\mathrm{logit}(t) - \mu\right)^2}{2\sigma^2}\right)
    \label{eq_logit_normal}
\end{align}

\noindent Since the curvature of the probability flow ODE solution is inconsistent at different times (see Fig.~\ref{fig:naive_flow_matching}). By mapping the time steps, reducing the time-step spacing where the curvature is large and enlarging the time step spacing where the curvature is slight can achieve better sampling quality. Since the time is sampled from the logit normal distribution during the training phase, the log-SNR $\lambda(t)$ in Eq.~\eqref{eq_snr} can be reparameterized with a new mean and standard deviation $\lambda(t, \mu, \sigma)$,

\begin{align}
    \lambda(t, \mu, \sigma) \equiv 2\,(\mu + \sigma\, \mathrm{logit}(t) )
    \;\;, \;\;\mathrm{logit}(t) \sim \mathcal{N}(0,1)
\end{align}

\noindent and the corresponding time-mapping relationship can be solved as

\begin{align} 
   \lambda(t, \mu, \sigma) = \lambda(t') 
   \Rightarrow \, t'(t, \mu, \sigma) = \frac{e^{\mu}}{e^{\mu}+(\frac{1}{t}-1)^{\sigma}}
\end{align}

\noindent The advantage of this time schedule is that one can flexibly adjust the mean and standard deviation to adapt to different sampling situations, and it is an effective means in practice.\cite{esser2024SD3}\cite{labs2025_flux1_kontext_flowmatching} In this work, the $(\mu, \sigma)$ are set to $(0, 1)$ during training phase and $(-0.3, 1)$ during inference phase.

All models in this work were trained using an identical hyperparameter configuration. We employed the AdamW optimizer~\cite{loshchilov2019decoupled_weight_decay} with the 1-cycle learning rate policy~\cite{smith2019superconvergence}, setting the maximum learning rate to $2 \times 10^{-4}$. Training was conducted with a batch size of 32, and we used 4 steps of gradient accumulation, resulting in an effective batch size of 128 for each parameter update.

The training and inference process is detailed in Algorithms~\ref{alg:fm_training_algorithm} and \ref{alg:fm_inference_algorithm}. The $\mathbf{x}_1$ is the spin configuration (spin angles) generated by the MCMC, $\mathbf{x}_0$ is the standard Gaussian noise with the same shape of spin configuration, $c_{\text{data}}$ is the corresponding temperature of the spin configuration and $p_t(t,\mu,\sigma)$ is the logit normal distribution in Eq.~\eqref{eq_logit_normal}. In the inference stage, since the probability flow ODE needs to be solved, the computational accuracy and time cost must be considered at the same time. On the premise that Flow Matching maintains the efficiency advantage over traditional algorithms, the 3rd-order Heun’s (Heun3) method within 10 steps is reliable. All Flow Matching sampling methods in this work use the 5-step Heun3 method to sample spin configurations.

\begin{algorithm}[H]
\caption{Flow Matching — Training}
\SetKwComment{Comment}{}{} 
\KwIn{Prior distribution $\mathcal{N}(0, \mathbb{I})$, time distribution $p(t,\mu,\sigma)$, data distribution $p_{\text{data}}$, conditioning data $c_{\text{data}}$, velocity field model $\mathbf{v}_\theta$, number of training steps $N$, batch size $B$, learning rate $\eta$}
\KwOut{Trained model parameters $\theta$}
\For{$n \gets 1$ \KwTo $N$}{
    Sample $(\mathbf{x}_1, c) \sim (p_{\text{data}}^B, c_{\text{data}}^B)$\;
    Sample $\mathbf{x}_0 \sim \mathcal{N}^B(0, \mathbb{I})$\;
    Sample $t \sim p^B(t,\mu,\sigma)$\;
    
    $\mathbf{x}_t \gets t\mathbf{x}_1 + (1-t)\mathbf{x}_0$ \Comment{Conditional flow path}
    $\mathbf{v}_\star \gets \mathbf{x}_1 - \mathbf{x}_0$ \Comment{Conditional vector field target}

    $\mathcal{L} \gets \frac{1}{B} \sum_{i=1}^{B} \bigl\|\mathbf{v}_\theta(\mathbf{x}_t^{(i)}, t^{(i)}, c^{(i)}) - \mathbf{v}_\star^{(i)}\bigr\|_2^2$\;
    $\theta \gets \theta - \eta\nabla_\theta\mathcal{L}$ \Comment{Update model parameters}
}
\KwRet{$\theta$}
\label{alg:fm_training_algorithm}
\end{algorithm}

\begin{algorithm}[H]
\caption{Flow Matching — Inference with Time-Mapping and 3rd-Order Heun's Method}
\SetKwComment{Comment}{}{} 
\KwIn{Prior distribution $\mathcal{N}(0, \mathbb{I})$, velocity field $\mathbf{v}_\theta$, number of steps $K$, conditioning $c_{\text{data}}$, time-mapping function $t'(\cdot, \mu, \sigma)$}
\KwOut{Sample $\mathbf{x}_K \approx p_{\text{data}}$}
Sample $\mathbf{x}_0 \gets \mathcal{N}(0, \mathbb{I})$\;
Generate uniform time steps $\tau_k = k/K$ for $k=0, \dots, K$\;
Map to non-uniform time steps $t_k = t'(\tau_k, \mu, \sigma)$ for $k=0, \dots, K$\;
\For{$k \gets 0$ \KwTo $K-1$}{
  $h \gets t_{k+1} - t_k$ \Comment{Non-uniform step size}
  
  $\mathbf{k}_1 \gets \mathbf{v}_\theta(\mathbf{x}_k, t_k, c_{\text{data}})$\;
  $\mathbf{k}_2 \gets \mathbf{v}_\theta(\mathbf{x}_k + \frac{h}{3}\mathbf{k}_1, t_k + \frac{h}{3}, c_{\text{data}})$\;
  $\mathbf{k}_3 \gets \mathbf{v}_\theta(\mathbf{x}_k + \frac{2h}{3}\mathbf{k}_2, t_k + \frac{2h}{3}, c_{\text{data}})$\;
  
  $\mathbf{x}_{k+1} \gets \mathbf{x}_k + \frac{h}{4}(\mathbf{k}_1 + 3\mathbf{k}_3)$\;
}
\KwRet{$\mathbf{x}_K$}  
\label{alg:fm_inference_algorithm}
\end{algorithm}

\section{More Physics Observables in Size Extrapolation}
\label{app:more_observables}

\begin{figure}[!ht]
    \centering
    \includegraphics[width=0.5\textwidth]{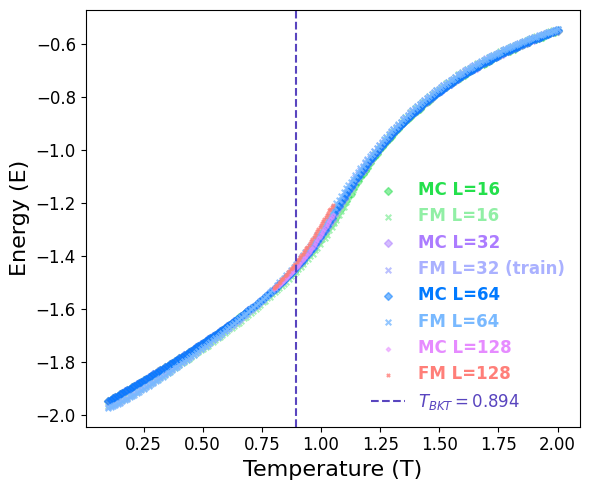}
    \caption{Size extrapolation of energy per site $E(T)$ for various system sizes ($L=16, 64, 128$). A single Flow Matching (FM) model, trained only on $32\times32$ data, generates configurations that show excellent agreement with the reference MCMC (MC) data. This highlights the model's ability to accurately extrapolate first-moment quantities.}
    \label{fig:energy_extrapolation}
\end{figure}

\begin{figure}[!ht]
    \centering
    \includegraphics[width=0.5\textwidth]{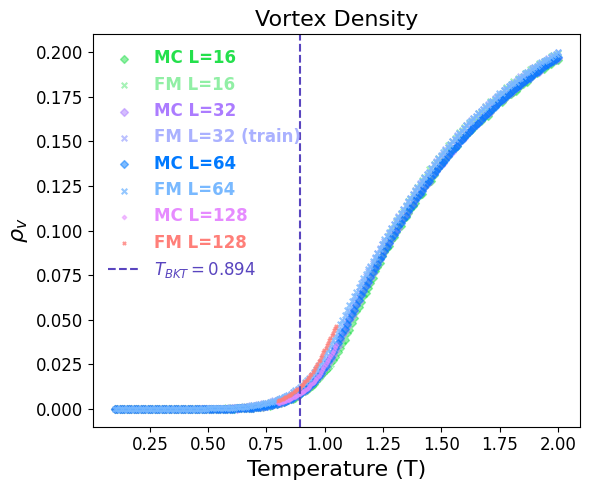}
    \caption{Size extrapolation of vortex density $\rho_v(T)$ for various system sizes ($L=16, 64, 128$). Generated by a single Flow Matching (FM) model trained only on $32\times32$ data, the results align well with the MCMC (MC) reference. This further validates the model's robust performance on observables not primarily driven by large-scale fluctuations.}
    \label{fig:vortex_density_extrapolation}
\end{figure}

\begin{figure}[!ht]
    \centering
    \includegraphics[width=0.5\textwidth]{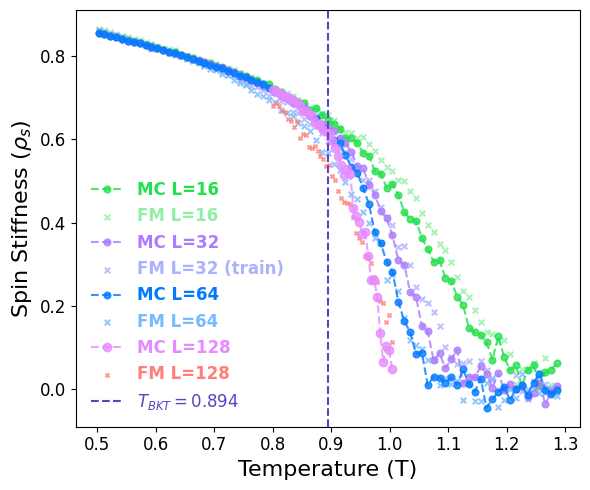}
    \caption{Size extrapolation of spin stiffness $\rho_s(T)$ for various system sizes ($L=16, 64, 128$), generated from a single FM model trained on $32\times32$ data. The statistical noise visible across all curves, including the MCMC (MC) reference, is partially attributable to the modest ensemble size of 1500 configurations per temperature used for both training and comparison. While the FM model captures the qualitative behavior, a rigorous finite-size scaling analysis for the BKT transition requires locating the intersection of the stiffness curve with the universal line $2T/\pi$, according to the Nelson-Kosterlitz criterion~\cite{PhysRevLett.39.1201}. As the figure shows, for extrapolated sizes ($L=64, 128$), a noticeable deviation between the FM and MC results appears near the critical transition temperature where this intersection is expected. This discrepancy invalidates a scaling analysis based on the intersection point and justifies our choice to use the more robust susceptibility peak (Sec.~\ref{sec:results_System_Sizes_Extrapolation}).}
    \label{fig:stiffness_extrapolation}
\end{figure}

To provide a comprehensive view of the model's extrapolation performance, this appendix presents a detailed comparison for several key observables across system sizes $L=16,\;32,\; 64,$ and $128$. The model was trained only on $L=32$ data with 1500 configurations per temperature from MCMC. The setup is identical to Sec.~\ref{sec:results_System_Sizes_Extrapolation}. The results distinguish between the model's robust performance on first-moment quantities (energy in Fig.~\ref{fig:energy_extrapolation}, vortex density in Fig.~\ref{fig:vortex_density_extrapolation}) and its challenges with the second-moment, fluctuation-related quantity (spin stiffness in Fig.~\ref{fig:stiffness_extrapolation}).

\section{Neural Network Architecture}
\label{app:nn}
Our model employs a U-Net architecture, a design renowned for its efficacy in image-to-image tasks due to its hierarchical structure that captures multi-scale features~\cite{ronneberger2015unet}. The network is composed of an encoder path, a bottleneck, and a decoder path, with skip connections linking corresponding levels of the encoder and decoder to preserve high-resolution spatial details. To handle the conditional nature of our generation task, the model integrates time and temperature embeddings, which are processed and fused before modulating the network's behavior at each block. 

\subsection{Core Building Blocks}
The fundamental component of our U-Net is the "ResnetBlock" (RB). Each block consists of two main sequential operations: a spatial processing layer and a feedforward neural network. The input to each operation is first normalized using Root Mean Square Layer Normalization, a variant of Layer Normalization known for its computational efficiency and performance~\cite{zhang2019RMSNorm}.

A key feature of the RB (see Fig.~\ref{fig:attention_res_block}) is its adaptive conditioning mechanism, inspired by the Adaptive Layer Normalization (adaLN) used in Diffusion Transformers~\cite{peebles2023DiT}. The conditional embedding, derived from time and temperature, is passed through a linear layer to generate shift and scale parameters. These parameters modulate the normalized feature maps within the block, allowing the network's behavior to be dynamically controlled by the input conditions.

While the U-Net's convolutional layers excel at learning local physical rules due to their inherent inductive biases of locality and translation equivariance, they are less effective at capturing the long-range correlations that emerge near critical points. To address this limitation, we replace the standard convolutional layers in RB with a specialized multi-head self-attention mechanism~\cite{vaswani2023attention_need} in the network's deeper, lower resolution levels. The structure is shown in Fig.~\ref{fig:attention_res_block}, which is denoted as "ResnetBlock (Attention)" (RBA) in the figure.

This attention mechanism incorporates a learnable 2D Rotary Position Embedding (RoPE), which has demonstrated a remarkable ability to handle multi-scale inputs and extrapolate across resolutions~\cite{heo2024RoPE2d}. By applying rotational transformations to the query and key vectors based on their spatial coordinates, 2D RoPE effectively encodes relative positional information directly into the attention mechanism. This allows the model to capture complex, long-range dependencies between spin sites across the entire lattice, a crucial capability for modeling systems at or near a phase transition where correlation lengths diverge. The frequencies of the rotary embeddings are treated as learnable parameters, enabling the model to adapt the positional encoding to the specific physics of the system.

\subsection{Optimization of Neural Network Architecture}
\begin{figure*}[t!]
    \centering
    \includegraphics[width=\textwidth]{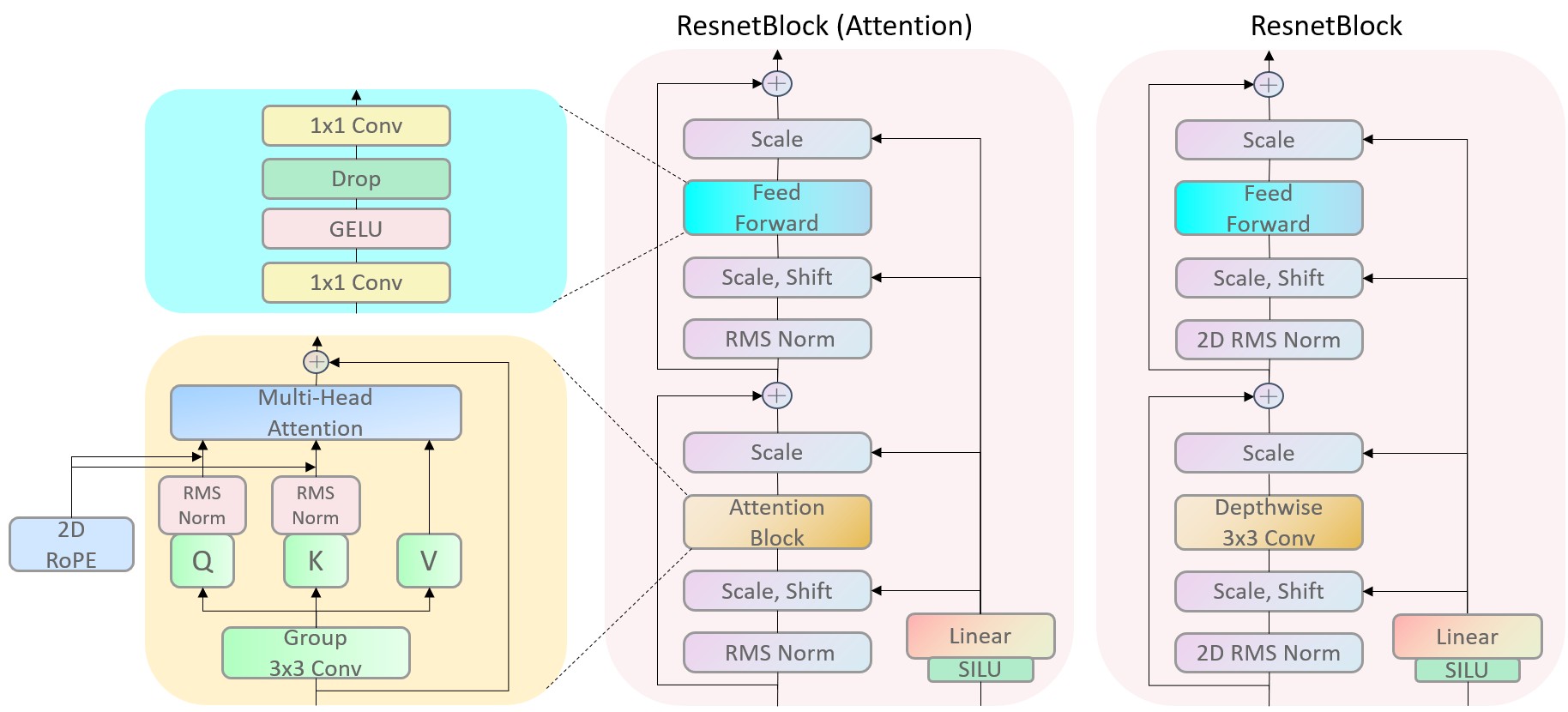}
    \caption{The core Building Blocks in U-Net. The left-hand side is "ResnetBlock (Attention)" (RBA), which is used in the network's deeper, lower resolution levels. "ResnetBlock" (RB) and "ResnetBlock (Attention)" (RBA) share the same structure except for the attention block. Please refer to the code in the following link for the detailed structure of the neural network. (\url{https://github.com/ToelUl/Flow-U-Net})
    }
    \label{fig:attention_res_block}
\end{figure*}

The entire model is optimized using "torch.compile" to enhance computational efficiency~\cite{pytorch}. For certain blocks ("ResnetBlock" and "ResnetBlock (Attention)" in Fig.~\ref{fig:attention_res_block}) in model is optimized using "epilogue fusion" and "max autotune" supported by PyTorch~\cite{pytorch} to perform operator fusion automatically. Please refer to the code in the following link for the detailed structure of the neural network. (\url{https://github.com/ToelUl/Flow-U-Net})

\section{Theoretical Receptive Field and Physical Implications}
\label{sec:receptive_field}

To rigorously understand the limitations of the model in capturing macroscopic phase coherence—specifically, the underestimation of spin stiffness upon extrapolation to larger sizes- we analyze the theoretical receptive field (RF) of our Flow Matching U-Net architecture. The RF determines the maximum spatial extent of input features that can influence a single spin at the output layer. 

For a convolutional neural network with a kernel size $K$, the theoretical receptive field at layer $l$ can be recursively calculated as \cite{araujo2019computing}:
\begin{equation}
    \text{RF}_{l} = \text{RF}_{l-1} + (K - 1) \times S_{\text{accum}}
\end{equation}
where $S_{\text{accum}}$ is the accumulated stride (or downsampling factor) from the input up to layer $l-1$.

Our U-Net consists of two downsampling levels implemented via pixel shuffle (spatial reduction by a factor of 2) and an identical upsampling strategy. The core components, including the RB (Fig.~\ref{fig:attention_res_block}) and the $QKV$ projection within the RBA (Fig.~\ref{fig:attention_res_block}) bottleneck, predominantly utilize $3 \times 3$ convolutions ($K=3$). The step-by-step RF expansion is detailed in Table~\ref{tab:receptive_field}.

\begin{table}[htbp]
    \centering
    \caption{Layer-by-layer calculation of the theoretical receptive field (RF) for the proposed U-Net architecture. $S_{\text{accum}}$ denotes the accumulated stride at the given stage. Please refer to the code in the following link for the detailed structure of the neural network. (\url{https://github.com/ToelUl/Flow-U-Net}) }
    \label{tab:receptive_field}
    \begin{tabular}{lccc}
        \hline\hline
        \textbf{Layer / Block} & \textbf{Kernel ($K$)} & $\mathbf{S_{\text{accum}}}$ & \textbf{Cumulative RF} \\
        \hline
        Input & - & 1 & 1 \\
        Initial Convolution & 3 & 1 & 3 \\
        Down Level 0 (Resnet + Downsample) & $3 \times 2$ & 1 & 7 \\
        \quad \textit{PixelUnshuffle} ($\times 2$) & - & 2 & 7 \\
        Down Level 1 (Resnet + Downsample) & $3 \times 2$ & 2 & 15 \\
        \quad \textit{PixelUnshuffle} ($\times 2$) & - & 4 & 15 \\
        Bottleneck (Attn QKV + Resnet) & $3 \times 2$ & 4 & 31 \\
        \quad \textit{PixelShuffle} ($\times \frac{1}{2}$) & - & 2 & 31 \\
        Up Level 1 (Resnet) & $3 \times 1$ & 2 & 35 \\
        \quad \textit{PixelShuffle} ($\times \frac{1}{2}$) & - & 1 & 35 \\
        Up Level 0 (Resnet) & $3 \times 1$ & 1 & 37 \\
        Final Convolution & 3 & 1 & \textbf{39} \\
        \hline\hline
    \end{tabular}
\end{table}

\noindent As shown in Table~\ref{tab:receptive_field}, the final structural receptive field of the U-Net is $39 \times 39$ pixels, corresponding to a theoretical receptive-field radius ($r_{\mathrm{RF}}$) of approximately 19 lattice sites. 

\begin{figure}[h]
    \centering
    \includegraphics[width=0.5\linewidth]{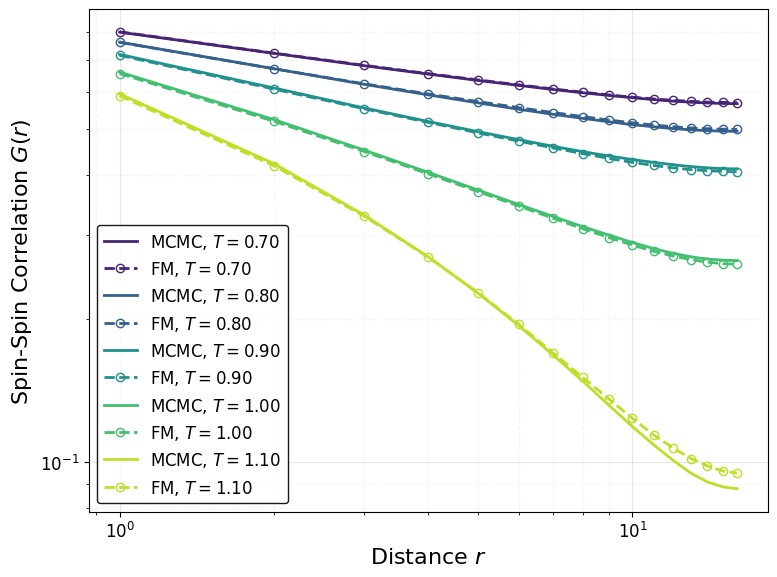} 
    \caption{
    Spin-spin correlation function $G(r)$ for the training lattice size $L=32$, plotted on a log-log scale. The Flow Matching (FM) generated configurations (dashed lines with markers) show strong overall agreement with the MCMC baseline (solid lines) across various temperatures, successfully capturing both the algebraic and exponential decay trends. A minor deviation at the maximum distance ($r \approx 16$) near the critical region ($T \approx 1.10$) is visible, reflecting the natural attenuation of the effective receptive field at its spatial limits.
    }
    \label{fig:corr_32}
\end{figure}

To empirically validate this spatial constraint, we compute the spin-spin correlation function $G(r)$ for both the training size ($L=32$) and an extrapolated size ($L=64$). As illustrated in Figure~\ref{fig:corr_32}, because the $39 \times 39$ structural RF envelops the $32 \times 32$ training system with periodic boundary conditions, the model generally behaves as a global operator. It successfully captures the overall trends of both the algebraic and exponential decays across most temperatures and distances. However, a slight deviation from the MCMC baseline is observable at the maximum independent distance ($r \approx 16$), particularly near the finite-size shifted critical temperature ($T=1.10$). This nuanced discrepancy aligns with our prior analysis: while the theoretical RF covers the entire lattice, the intrinsic Gaussian decay of the effective receptive field (ERF) towards its boundaries slightly attenuates the model's capacity to resolve the most extreme system-wide critical correlation perfectly.

\begin{figure}[h]
    \centering
    \includegraphics[width=0.5\linewidth]{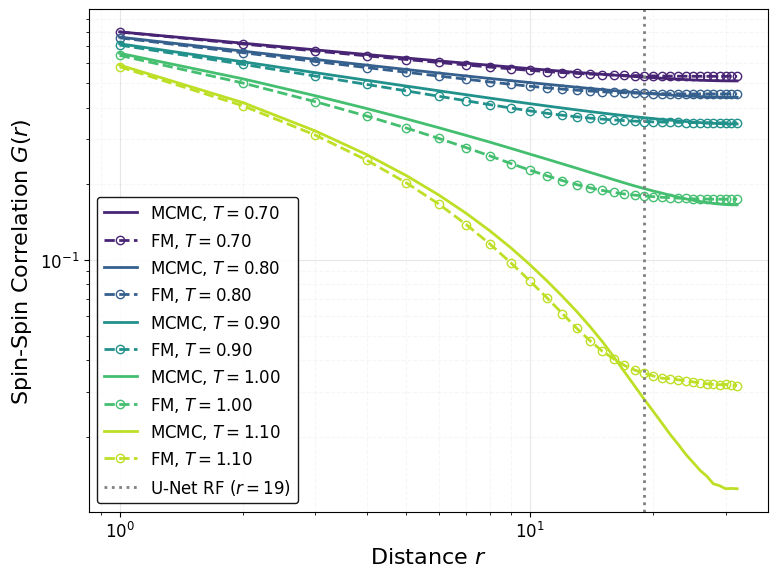} 
    \caption{
    Log-log plot of the spin-spin correlation function $G(r)$ extrapolated to a $64 \times 64$ lattice. The vertical dotted line indicates the theoretical U-Net receptive-field radius ($r_{\mathrm{RF}}\approx 19$). Noticeable deviations from the MCMC baseline occur beyond this boundary, particularly near the critical region ($T \approx 1.1$), illustrating the fundamental structural limits of the model in capturing long-distance phase coherence across the full lattice.}
    
    \label{fig:corr_64}
\end{figure}

However, the spatial constraints of the network manifest as a fundamental physical cutoff when extrapolating zero-shot to larger lattices. Figure~\ref{fig:corr_64} demonstrates this phenomenon for a $64 \times 64$ lattice. For short and intermediate distances ($r < r_{\mathrm{RF}} \approx 19$), the FM-generated configurations maintain strong agreement with MCMC. Conversely, visible deviations emerge for $r > 19$, particularly at temperatures near the finite-size shifted critical region ($T \approx 1.1$), where critical fluctuations are most pronounced. 

This deviation is consistent with general convolutional-network theory. As shown by Luo et al.~\cite{luo2017understandingeffectivereceptivefield}, the \textit{effective} receptive field (ERF) of a deep CNN typically occupies only a fraction of the theoretical RF. It is concentrated near its center, with a Gaussian-like decay toward the boundaries. This provides a natural explanation for why the present model reproduces short- and intermediate-range correlations more faithfully than it does the longest-distance correlations on extrapolated lattices.

In addition, the self-attention bottleneck uses Rotary Position Embedding (RoPE)~\cite{su2023roformerenhancedtransformerrotary} to encode relative positional information. Although we do not isolate this effect through a dedicated ablation study in the present work, previous studies on positional or length extrapolation~\cite{chen2023extendingcontextwindowlarge, peng2026yarnefficientcontextwindow} suggest that RoPE-based attention can become less reliable when evaluated at relative distances not seen during training. In our setting, this may provide an additional source of difficulty for transferring long-range correlations from the $32\times32$ training lattice to larger unseen lattices.

We therefore interpret the observed loss of long-range phase coherence primarily as a consequence of the finite spatial range effectively accessible to the compact architecture, with possible additional degradation from out-of-distribution positional extrapolation in the attention bottleneck. This interpretation is fully consistent with the correlation-function results in Figs.~\ref{fig:corr_32} and \ref{fig:corr_64}: the model remains accurate at short and intermediate distances, but progressively deviates at longer distances on extrapolated lattices. Consequently, while the framework is computationally efficient at generating physically aligned warm-start configurations, rigorous recovery of macroscopic phase coherence---especially for observables such as spin stiffness---still requires subsequent MCMC refinement.

\section{Comparison between U-Net and DiT Architectures}
\label{app:unet_vs_dit_comparison}

To show that the inductive biases embedded in CNN architectures of U-Net are the key point for training on data of a fixed system size and extrapolating to other sizes, we compare the results using another robust neural network architecture: the Diffusion Transformer~\cite{peebles2023DiT}. 

Vision Transformers~\cite{vit} and, by extension, the Diffusion Transformer (DiT)~\cite{peebles2023DiT} variants recently adopted for generative modeling, rely on minimal inductive biases. Transformer's core is the self-attention mechanism~\cite{vaswani2023attention_need}. This mechanism empowers them with global receptive fields and permutation equivariance, enabling each image patch (or token) to interact with all others regardless of spatial proximity. The only explicit inductive bias is introduced through the positional encoding scheme, which provides spatial context but does not enforce strict locality. As a consequence, Transformers excel when long-range correlations or nonlocal dependencies dominate the target distribution, conditions frequently encountered in natural language, graphs, and physical systems near criticality where correlation lengths diverge. But in our case, the underlying data-generating process is governed by a physical Hamiltonian with strictly local interactions.

\begin{figure}[!ht]
    \centering
    \includegraphics[width=0.5\textwidth]{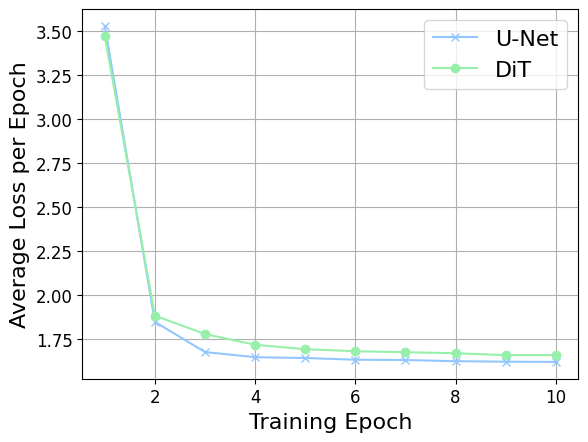}
    \caption{Training loss curves of U-Net and DiT, where each point represents the average loss per epoch. The two models exhibit similar convergence behavior, but U-Net performs better. Inference sampling results for the two models are shown in Figs.~\ref{fig:dit_vs_unet_training_size_32} and ~\ref{fig:dit_vs_unet_testing_size_64}.} 
    \label{fig:loss_comparison}
\end{figure}

We conducted a comparative experiment under identical training conditions to validate this hypothesis. Both the U-Net and DiT models were trained on the same dataset from MCMC simulations of a $32\times32$ system. Notably, to specifically accommodate the potentially higher data requirements of the larger DiT model, we increased the number of training samples to $2500$ configurations per temperature point (up from 1500 used in the system size extrapolation experiment, see Sec.~\ref{sec:results_System_Sizes_Extrapolation}). Both models were trained for the same number of epochs with an identical optimizer and learning rate schedule, ensuring a controlled comparison. Both models ensure that the training has converged and the training loss curve is shown in the Fig.~\ref{fig:loss_comparison}.

A potential concern in this comparison is the significant disparity in model size (Table~\ref{tab:model_comparison}), where the DiT model has nearly six times more parameters than the U-Net. One might argue that the DiT's poorer performance could stem from a larger, more complex model that is inherently more difficult to train, rather than from a fundamental misalignment of its inductive biases.

\begin{table}[h!]
\caption{Comparison of U-Net and DiT model architectures. Due to its structural limitations, the DiT model is significantly larger regarding parameters and computational cost (floating point operations, FLOPs). All training and inference conditions were held constant between the two models.}
\centering

\begin{tabular*}{\columnwidth}{@{\extracolsep{\fill}} l c c}
\hline
Metric          & U-Net     & DiT         \\
\hline
Parameters      & 214.65 K  & 1262.98 K   \\
FLOPs           & 0.9776 G  & 6.4847 G    \\
Memory (params) & 0.82 MB   & 4.82 MB     \\
\hline
\end{tabular*}
\label{tab:model_comparison}
\end{table}

However, we argue that this disparity in parameters is not a confounding factor but a direct consequence of the architectural differences and is, in fact, central to our conclusion. The DiT architecture, lacking a built-in locality bias, requires a much larger parameter space to learn local physical relations from data—a task for which the U-Net is purpose-built via its convolutional structure. The DiT's substantial parameter count is a necessity from its lack of correct inductive bias. Therefore, its struggle to train effectively under these conditions is not an artifact of an unfair comparison but rather a direct demonstration of our central thesis: for physical systems governed by local interactions, having the correct architectural inductive bias is far more critical than raw model capacity. The training challenge itself highlights the inherent efficiency and superiority of the U-Net's architecture for this class of problems.

\begin{figure*}[t!]
    \centering
    \includegraphics[width=\textwidth]{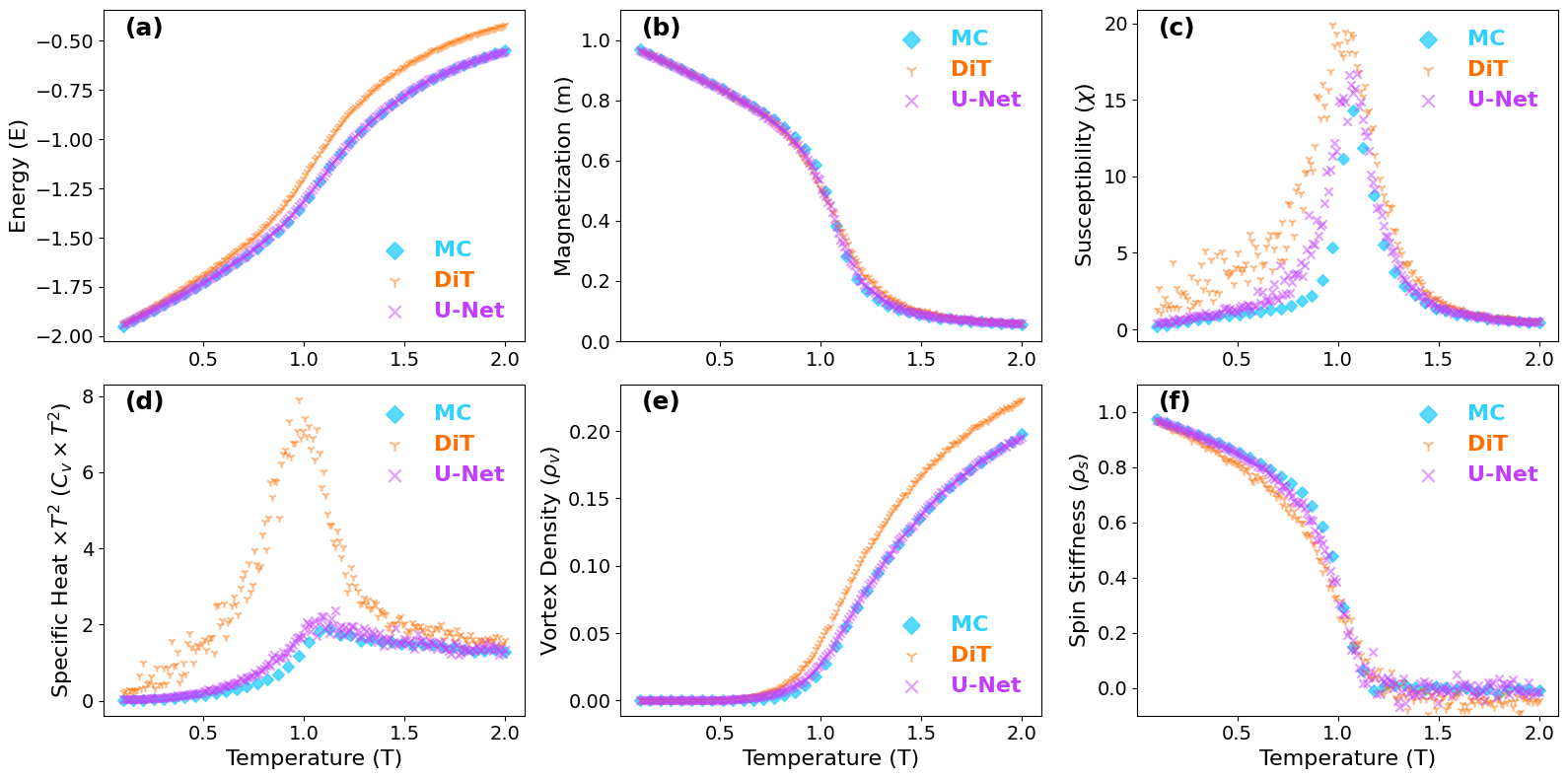}
    \caption{This figure shows the performance of the U-Net and DiT trained and sampled at the same size ($32\times 32$). The MCMC (MC) data used for training and comparison have a temperature interval 0.05. The sampling temperature interval of U-Net and DiT after training with Flow Matching is 0.01. The figure clearly shows that even when sampling at the training size, DiT, which lacks the correct inductive bias, cannot learn the correct physical behavior with limited training data like the U-Net.}
    \label{fig:dit_vs_unet_training_size_32}
\end{figure*}

\begin{figure*}[t!]
    \centering
    \includegraphics[width=\textwidth]{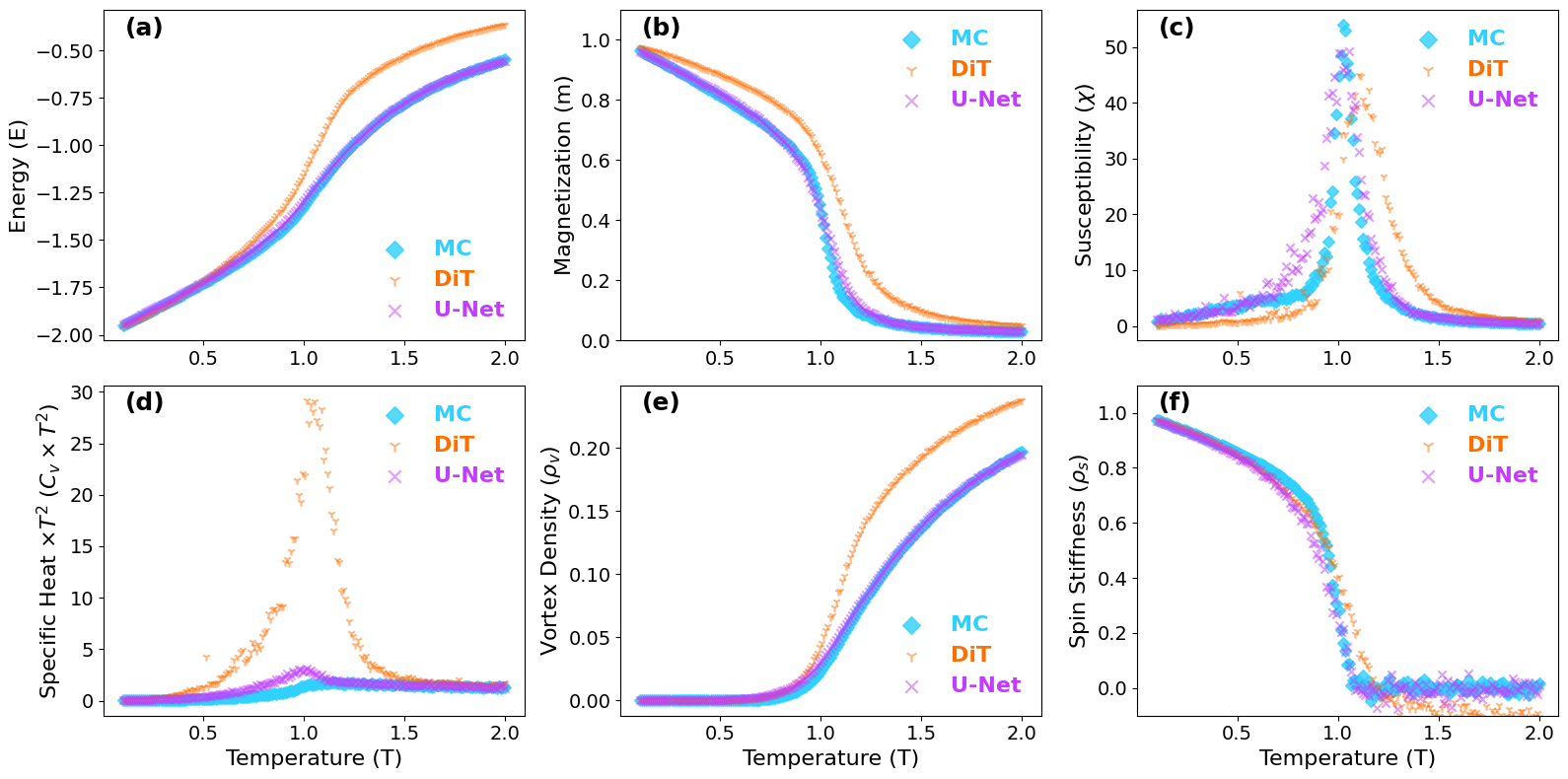}
    \caption{This figure shows the performance of U-Net and DiT trained at size $32\times 32$ and sampled at a larger size $64\times 64$. The MCMC data used for training at size $32\times 32$ had a temperature interval of 0.05. U-Net and DiT, after training with Flow Matching, had a sample temperature interval of 0.01. The MCMC (MC) data used for comparison at the larger size $64\times 64$ in the figure had a temperature interval of 0.01. The figure shows that DiT, which lacks the correct inductive bias, has worse results under size extrapolation (compared to Figs.~\ref{fig:dit_vs_unet_training_size_32}) and does not exhibit correct finite-size scaling behavior.}
    \label{fig:dit_vs_unet_testing_size_64}
\end{figure*}

The DiT architecture, lacking strong locality bias, must deduce the principle of locality entirely from the training data. Its self-attention mechanism initially treats all pairs of sites as potentially relevant, placing a significant learning burden on the model to discover that the system's dynamics are overwhelmingly dominated by nearest-neighbor interactions. This leads to a critical challenge in generalization, as observed in our Flow Matching experiments (see Figs.~\ref{fig:dit_vs_unet_training_size_32} and~\ref{fig:dit_vs_unet_testing_size_64}). The DiT model tends to learn correlations and patterns that are specific to the fixed system size seen during training, rather than the universal, size-invariant rule of local physical law. When extrapolating to larger, unseen system sizes ($64\times 64$), the global attention patterns learned in the smaller system are no longer directly applicable, leading to a significant performance degradation.

Conversely, the powerful inductive biases of a CNN-based architecture in the U-Net align perfectly with the fundamental structure of the local Hamiltonian. The convolutional kernel intrinsically learns a local rule that is, by its very nature, independent of the total system size. This inherent understanding of locality and translation equivariance means the learned physical principles are more robust and readily generalizable. The network does not need to waste capacity learning the concept of locality; it can instead focus on modeling the specific nature of these local interactions. This alignment explains why the U-Net architecture demonstrates superior extrapolation capabilities, successfully generalizing to different system sizes after being trained on only one. 

This observation is consistent with a large body of literature in computer vision comparing Transformers and CNNs. The seminal work on Vision Transformers by Dosovitskiy et al.~\cite{vit} explicitly states that Transformers lack the inductive biases of locality and translation equivariance, making them heavily reliant on massive-scale pre-training to generalize effectively. Subsequent analyses, such as the work by Raghu et al.~\cite{vit_vs_cnn}, have provided further evidence, showing that ViTs must learn to represent local information from scratch, a task for which CNNs are architecturally predisposed. Our comparative experiment with the DiT model, therefore, serves as a direct demonstration of this principle within a physical simulation context: the model with the correct inductive bias (U-Net) not only performs better but does so with a fraction of the parameters, because its architecture is already aligned with the fundamental, local nature of the physical system.

\section{GPU-accelerated Monte Carlo Method}
\label{app:mc_gpu}

We employ a highly optimized MCMC simulation framework implemented in PyTorch~\cite{pytorch}. While the Wolff~\cite{PhysRevLett_Wolff} or Swendsen–Wang~\cite{PhysRevLett.58.86} cluster algorithm is known to be more efficient in reducing autocorrelation steps for the 2D XY model, its sequential cluster-building process is not amenable to efficient GPU parallelization. Our choice of the Metropolis-Hastings algorithm with a checkerboard update scheme is motivated by its exceptional performance on massively parallel architectures, making it a powerful and relevant baseline for modern hardware. The full implementation of our algorithm is available on GitHub: (\url{https://github.com/ToelUl/Thermal-Phase-Transition-MC-GPU-Simulation}). The algorithm is designed to leverage the massively parallel architecture of modern Graphics Processing Units (GPUs), enabling large-scale simulations with high statistical precision. The core design incorporates three key parallelization strategies: a checkerboard update scheme for intra-lattice parallelism, the simulation of multiple independent Markov chains, and the replica-exchange method (also known as Parallel Tempering~\cite{Hukushima_1996}) across a range of temperatures.

\subsection{Parallel Architecture and Data Layout}

The fundamental data structure for the spin configurations is a four-dimensional tensor of shape $B \times C \times L \times L$, where:
\begin{itemize}
    \item $B$ is the batch size, representing the number of different temperatures simulated simultaneously.
    \item $C$ is the number of independent Monte Carlo chains for each temperature.
    \item $L$ is the linear size of the square lattice.
\end{itemize}
This layout allows for all temperatures and chains to be processed in parallel. Operations are vectorized across the first two dimensions ($B$ and $C$), which maps naturally to the single instruction, multiple data (SIMD) paradigm of GPU computing. The use of $C$ independent chains per temperature not only accelerates the accumulation of statistics but also enhances the statistical independence of the collected samples, providing a more robust ensemble average.

\subsection{Checkerboard Metropolis-Hastings Update}

A standard single-spin flip Metropolis-Hastings algorithm introduces data dependencies, as the update of a spin requires knowledge of its neighbors, precluding simultaneous updates of adjacent spins. To overcome this limitation and maximize parallel efficiency, we implement a \textbf{checkerboard update scheme}~\cite{Weigel_2011}. The lattice is decomposed into two independent sub-lattices, 'A' and 'B' (analogous to the black and white squares of a checkerboard).

A single Monte Carlo sweep consists of two sequential steps:
\begin{enumerate}
    \item All spins on sub-lattice A are updated simultaneously. Since no spin on sub-lattice A is a nearest neighbor to any other spin on A, these updates are fully independent and can be performed in a single, vectorized operation on the GPU.
    \item All spins on sub-lattice B are then updated simultaneously in the same manner.
\end{enumerate}

For the XY model, where the Hamiltonian is given by:
\begin{align}
    H = -J \sum_{\langle i,j \rangle} \cos(\theta_i - \theta_j)
\end{align}
the local update involves proposing a new angle $\theta'_i = \theta_i + \delta\theta$ for a spin $i$, where $\delta\theta$ is a random perturbation. The change in energy, $\Delta E$, is computed based only on the local spin and its neighbors. The proposed move is then accepted with the probability:
\begin{align}
    P_{\mathrm{acc}}(\theta_i \to \theta'_i) = \min\left(1, e^{-\Delta E / T}\right)
\end{align}
where TT is the temperature of the specific replica. This entire process, from proposing moves to accepting them, is executed in parallel for all spins on the active sub-lattice across all chains and temperatures.

\subsection{Parallel Tempering (Replica-Exchange)}

To mitigate critical slowing down near phase transitions and to ensure efficient exploration of the configuration space, our framework implements the \textbf{Parallel Tempering (or replica-exchange) algorithm}~\cite{Hukushima_1996}. This method involves running simulations at multiple temperatures $T_1, T_2, \dots, T_B$ in parallel. Periodically, an exchange of the entire spin configurations between two adjacent replicas, say at temperatures $T_i$ and $T_{i+1}$ with energies $E_i$ and $E_{i+1}$, is proposed.

The exchange is accepted with a probability that satisfies the detailed balance condition:

\begin{align}
    P(\mathrm{swap}) = \min\left(1, \exp\left[ \left(\frac{1}{T_i} - \frac{1}{T_{i+1}}\right) (E_i - E_{i+1}) \right]\right)
\end{align}

This allows configurations from high-temperature replicas, which explore the energy landscape more broadly, to diffuse to lower temperatures, preventing them from getting trapped in local energy minima. To ensure detailed balance over the entire system, exchange attempts are alternated between pairs $(T_i, T_{i+1})$ with even $i$ and those with odd $i$. This technique significantly improves the convergence and ergodicity of the simulation, especially for complex systems.

By combining these three levels of parallelism, the algorithm achieves a substantial reduction in simulation time while generating statistically robust and independent samples across a wide range of temperatures.

\section{Information about Computing Resources.}
\label{app:computational_details}
All numerical experiments, including model training, inference, and Monte Carlo simulations, were conducted on a workstation with the following specifications. 

\begin{itemize}
    \item \textbf{CPU:} AMD Ryzen 9 9950X 16-Core Processor
    \item \textbf{GPU:} 1x NVIDIA GeForce RTX 4080 (16 GB VRAM)
    \item \textbf{Operating System:} Windows Subsystem for Linux 2 (WSL2)
    \item \textbf{Core Libraries~\cite{pytorch, cudnn,lipman2024flow_matching_guide_code}:}
    \begin{itemize}
        \item PyTorch: 2.7.1 (+cu128)
        \item CUDA Toolkit (used by PyTorch): 12.8
        \item cuDNN: 9.7.1
        \item flow\_matching: 1.0.10
    \end{itemize}
\end{itemize}

\bibliographystyle{iopart-num}  
\bibliography{ref}

\end{document}